\documentclass [preprint]{aastex}

\lefthead{Taniguchi et al.}
\righthead{SUBARU OBSERVATIONS OF THE HST COSMOS FIELD}

\begin{document}


\title{THE COSMIC EVOLUTION SURVEY (COSMOS):
       SUBARU OBSERVATIONS OF THE HST COSMOS FIELD
       \altaffilmark{1}}

\author{Y. Taniguchi  \altaffilmark{2},
        N. Scoville         \altaffilmark{3,4},
        T. Murayama         \altaffilmark{5},
        D. B. Sanders       \altaffilmark{6},
        B. Mobasher         \altaffilmark{7},
        H. Aussel           \altaffilmark{8},
        P. Capak            \altaffilmark{3},
        M. Ajiki            \altaffilmark{5},
        S. Miyazaki         \altaffilmark{9},
        Y. Komiyama         \altaffilmark{10},
        Y. Shioya           \altaffilmark{2},
        T. Nagao            \altaffilmark{10, 11},
        S. S. Sasaki        \altaffilmark{2, 3, 5},
        J. Koda             \altaffilmark{3},
        C. Carilli          \altaffilmark{12},
        M. Giavalisco       \altaffilmark{13},
        L. Guzzo            \altaffilmark{14},
        G. Hasinger         \altaffilmark{15},
        C. Impey            \altaffilmark{16},
        O. LeFevre          \altaffilmark{17},
        S. Lilly            \altaffilmark{18},
        A. Renzini          \altaffilmark{19},
        M. Rich             \altaffilmark{20},
        E. Schinnerer       \altaffilmark{21},
        P. Shopbell         \altaffilmark{3},
        N. Kaifu            \altaffilmark{10},
        H. Karoji           \altaffilmark{10},
        N. Arimoto          \altaffilmark{10},
        S. Okamura          \altaffilmark{22}, and
        K. Ohta             \altaffilmark{23}
}

\altaffiltext{1}{Based on data collected at 
	Subaru Telescope, which is operated by 
	the National Astronomical Observatory of Japan.}
\altaffiltext{2}{Physics Department, Graduate School of Science \& 
    Engineering, Ehime University, 2-5 Bunkyo-cho, Matsuyama
    790-8577, Japan}
\altaffiltext{3}{California Institute of Technology, MC 105-24, 1200 East
        California Boulevard, Pasadena, CA 91125}
\altaffiltext{4}{Visiting Astronomer, University of Hawaii, 2680 Woodlawn
        Drive,  Honolulu, HI, 96822}
\altaffiltext{5}{Astronomical Institute, Graduate School of Science,
        Tohoku University, Aramaki, Aoba, Sendai 980-8578, Japan}
\altaffiltext{6}{Institute for Astronomy, 2680 Woodlawn Drive, 
        University of Hawaii, Honolulu, HI, 96822}
\altaffiltext{7}{Space Telescope Science Institute, 3700 San Martin
        Drive, Baltimore, MD 21218}
\altaffiltext{8}{Service d'Astrophysique, CEA/Saclay, 91191 
        Gif-sur-Yvette, France}
\altaffiltext{9}{Subaru Telescope, National Astronomical Observatory,
          650 N. A'ohoku Place, Hilo, HI 96720, USA}
\altaffiltext{10}{National Astronomical Observatory of Japan,
          2-21-1 Osawa, Mitaka, Tokyo 181-8588, Japan}
\altaffiltext{11}{INAF -- Osservatorio Astrofisico di Arcetri, 
        Largo Enrico Fermi 5, 50125 Firenze, Italy}
\altaffiltext{12}{National Radio Astronomy Observatory, 
    P.O. Box 0, Socorro, NM 87801-0387}
\altaffiltext{13}{Space Telescope Science Institute, 3700 
    San Martin Drive, Baltimore, MD 21218}
\altaffiltext{14}{Osservatorio Astronomico di Brera, via Brera,
        Milan, Italy}
\altaffiltext{15}{Max Planck Institut fuer Extraterrestrische 
    Physik,  D-85478 Garching, Germany}
\altaffiltext{16}{Steward Observatory, University of Arizona, 
    933 North Cherry Avenue, Tucson, AZ 85721}
\altaffiltext{17}{Laboratoire d'Astrophysique de Marseille, 
    BP 8, Traverse du Siphon, 13376 Marseille Cedex 12, France}
\altaffiltext{18}{Department of Physics, Swiss Federal Institute 
    of Technology (ETH-Zurich), CH-8093 Zurich, Switzerland}
\altaffiltext{19}{European Southern Observatory,
    Karl-Schwarzschild-Str. 2, D-85748 Garching, Germany}
\altaffiltext{20}{Department of Physics and Astronomy, 
    University of California, Los Angeles, CA 90095}
\altaffiltext{21}{Max Planck Institut f\"ur Astronomie, 
    K\"onigstuhl 17, Heidelberg, D-69117, Germany}
\altaffiltext{22}{Department of Astronomy, Graduate School of Science,
    The University of Tokyo, 7-3-1 Hongo, Bunkyo-ku, Tokyo 113-0033, Japan}
\altaffiltext{23}{Department of Astronomy, Graduate School of Science,
    Kyoto University, Kitashirakawa, Sakyo-ku, Kyoto 606-8502, Japan}

\begin{abstract}
We present deep optical imaging observations of 2 square degree area, covered
by the Cosmic Evolution Survey (COSMOS), made by the prime-focus Camera
(Supreme-Cam) on the 8.2m Subaru Telescope. Observations were done in six
broad-band [$B$ (4459.7 \AA), $g^\prime$ (4723.1 \AA), $V$ (5483.8 \AA), 
$r^\prime$ (6213.0 \AA), $i^\prime$ (7640.8 \AA), $z^\prime$ (8855.0 \AA)],
and one narrow-band ($NB816$) filters. A total
of $10^6$ galaxies were detected to $i'\sim 26.5$ mag. These data, combined
with observations at $u^*$ and $K$-band are used to construct the photometric
catalogs for the COSMOS and to measure their photometric redshifts, multi-band
spectral energy distributions, stellar masses and identification of
high redshift candidates. This catalog provides multi-waveband data for 
scientific analysis of the COSMOS survey.
\end{abstract}


\section{INTRODUCTION}

The Cosmic Evolution Survey (COSMOS) is a treasury program on
the Hubble Space Telescope (HST), awarded a total of 640 HST orbits,
carried out in two cycles (320 orbits in cycles 12 and 13 each;
Scoville et al. 2006a; Koekemoer et al. 2006).
COSMOS is a 2 square degree imaging survey of an equatorial field in
 $I_{\rm 814}$ band, using the Advanced Camera for Surveys (ACS). The HST ACS
observations provide high resolution imaging to map the morphology
of galaxies as a function of environment and epoch covering from
high redshift ($z \sim 6$) to the nearby ($z \sim 0$) universe.
Since substantial large-scale structures (e.g., voids, filaments, groups and
clusters of galaxies) occur on scales up to 100 Mpc in the comoving frame, 
our 2 square degree COSMOS field can adequately map galaxy evolution over the
full range of environments\footnote{Note that 1.4 degree corresponds to
a co-moving scale of  46.1 Mpc for $z=0.5$,
80.7 Mpc for $z=1$,
126.6 Mpc for $z=2$,
155.3 Mpc for $z=3$,
175.2 Mpc for $z=4$,
190.0 Mpc for $z=5$, and
201.5 Mpc for $z=6$,
under a flat cosmology of 
$\Omega_\Lambda = 0.7$,  $\Omega_{\rm m} = 0.3$, and
$H_0= 70$ km s$^{-1}$ Mpc$^{-1}$.}.
It is also interesting to note that our survey volume at high redshift 
is similar to that of the Sloan Digital Sky Survey (York et al. 2000)
at low redshift.

Our ACS survey depth, $I_{\rm 814, lim} \simeq 27$ AB allows us to
detect of the order of $\sim 10^6$ sources in the COSMOS area. 
Therefore, the COSMOS project is fundamental to virtually all areas of 
galaxy evolution, identification of different classes of objects, 
evolution of large-scale structures as well as that of dark matter.
In particular, the following interesting issues can be covered by our 
COSMOS project  using statistically large samples:
[1] the evolution of galaxies, clusters, large-scale structure, and
 cold dark matter on mass scales up to $> 10^{14} M_\odot$ as a function of redshift,
[2] the formation, assembly, and evolution of galaxies as a function of 
 large-scale structure environment, morphology and redshift,
[3] the cosmic star formation history as a function of 
 large-scale structure environment, morphology and redshift, and
[4] detailed study of the nature, morphology and clustering properties
of different populations of galaxies such as active galactic nuclei (AGNs), 
extremely red objects, Ly$\alpha$ emitters (LAEs), Lyman Break Galaxies (LBGs), 
and star-forming galaxies, and their evolution with redshift.

However, to understand the whole evolution of galaxies, AGNs, and dark matter,
it is absolutely necessary to obtain multi-wavelength observations with high spatial 
resolution from X-ray to radio. In particular, we need optical multi-band
images of the COSMOS field, as we only take $I_{\rm 814}$ band data with ACS. 
Therefore, ground-based optical observations are also an essential part of 
the COSMOS project. Such data will be helpful in studying stellar contents
of one million galaxies at various redshifts and in estimating photometric 
redshifts with reasonable accuracy.
This could best be accomplished by using wide-area CCD detectors on 8m 
class telescopes.  In this respect, the Subaru Prime-Focus Camera, 
Suprime-Cam (Miyazaki et al. 2002), on the Subaru Telescope 
(Kaifu et al. 2000; Iye et al. 2004), provides its superior imaging capability
because of its very wide field of view ($34^\prime \times 27^\prime$). 

During the period January 2004 to April 2005, we obtained deep
optical images of the COSMOS field with Suprime-Cam with the following seven 
filters; $B$, $g^\prime$, $V$, $r^\prime$,  $i^\prime$,  $z^\prime$, and $NB816$.
We describe these observations in detail; see also Taniguchi  et al. (2005). 
The last filter is the narrowband filter centered at 815 nm with a FWHM of 12 nm.
The broad-band data will be used to measure photometric redshifts and
stellar mass (Mobasher et al. 2006), identify large scale structures
(Scoville et al. 2006b; Guzzo et al. 2006), estimate local densities 
(Capak et al. 2006b) and optically identify sources detected in X-ray (Hasinger et al. 2006), 
radio (Schinnerer et al. 2006), and infrared (Sanders et al 2006) wavelengths. 
Combined with the narrow-band (NB816) observations, these will be used to identify
LAEs at $z=5.7$ (Murayama et al 2006), [O {\sc ii}] emitters at $z=1.2$ (Takahashi et al.
2006), and H$\alpha$ emitters at $z=0.24$ (Shioya et al. 2006).

In the present paper we describe in detail the observational procedures, 
filters, sensitivities and completeness of the COSMOS Subaru Suprime-Cam 
imaging (see also Taniguchi et al. 2005).
Science investigations and their initial results are presented in the
aforementioned papers.
Throughout this paper, we use the AB magnitude system.

\section{OBSERVATIONS}

\subsection{Observational Strategy}

The COSMOS field covers an area of  1.4 $\times$ 1.4 degree,
centered at RA(J2000) = 10:00:28.6 and DEC(J2000) = +02:12:21.0.
The Suprime-Cam consists of ten $2048\times 4096$ CCD chips and
provides a very wide field of view, $ 34^\prime \times 27^\prime$ in
$10240 \times 8192$ pixels
(0\farcs202 pixel$^{-1}$).
Despite the large field of view of the Supreme-Cam, we need a total of
nine pointings to cover the whole COSMOS area. This requires special mapping
(i.e. dithering) patterns to carry out the imaging observations. In order 
to obtain accurate astrometry, we also need to arrange the patterns 
to overlap.
Furthermore, we need to take care of gaps ($3^{\prime\prime}$--$4^{\prime\prime}$
or  $16^{\prime\prime}$--$17^{\prime\prime}$)
between the CCD chips of the Suprime-Cam.
In order to save observing time, we need an efficient mapping
pattern with minimum pointings for covering the whole field of COSMOS
with minimizing the shallower edges around the outside of the mosaic.
On one hand, in order to achieve reliable astrometry covering
the whole COSMOS field, it is also necessary to have a half-array shifted
data set.

Taking these two points into account, we use the following two
mapping patterns; Pattern A and Pattern C. Originally, we had another 
Pattern B. However, we did not use this in our observations. From this historical
reason, we refer our two mapping patterns as Pattern A and C
throughout this paper.
Pattern A is a half-array shifted mapping method in which 12 pointings
are necessary to map the whole COSMOS field; see Figure 1. The detailed
dithering properties are given in Table 1.
Pattern C is our most efficient mapping method in which only nine pointings
are enough to map the whole field; see Figure 2. The detailed
dithering properties are given in Table 2.
Pattern A was used with a relatively short unit exposure time
(e.g., a few to several minutes) because this pattern data were
used to obtain better astrometry and photometry in the whole COSMOS field
(Aussel et al. 2006; Capak et al. 2006a).
On the other hand, Pattern C was used with a longer unit integration
time because this pattern was used to obtain deeper data efficiently.  

\subsection{Observational Programs and Runs}

Our Suprime-Cam observations of the COSMOS field have been made during a
period between 2004 January and 2005 March, consisting of
three common-use observing programs.
Four nights were allocated within the University of Hawaii observing time
on the Subaru Telescope during a period between 2004 Feb and 2004 March; 
PI = N. Scoville.
A summary of the observational programs is given in Table 3.
It is noted that the two programs, S03B-239 and S04B-142, were
allocated as an Open Use Intensive Program; such Intensive Programs 
provide the opportunity for researchers to proceed with large programs of
advanced study that can only be achieved with the unique capability of 
Subaru Telescope and its instruments and needs an allocation of significant
telescope time. 
Another Intensive Program (COSMOS-21) was also accepted in the semester S05B
(S05B-013); note that ^^ 21" means 21 filters in the optical window.
 However, details of observations of S05B-013 will be given 
in a forthcoming paper.

Including the University of Hawaii time, 24.5 nights were allocated 
in total for our Suprime-Cam imaging of the COSMOS field.
These observations were carried out in eight observing runs; 
see Table 4.
During these runs, we obtained optical images of the COSMOS field
with the Johnson broad band filters, $B$ and $V$, the SDSS
broad band filters, $g^\prime$,
$r^\prime$, $i^\prime$, and  $z^\prime$, and a narrowband
filter, $NB816$\footnote{Our SDSS broad-band filters are designated as
$g^+$, $r^+$, $i^+$, and $z^+$ in Capak et al.
(2006a) to distinguish from the original SDSS filters. Also, our $B$ and $V$ 
filters are designated as $B_J$ and $V_J$ in Capak et al. (2006) where
$J$ means Johnson and Cusins filter system used in Landolt (1992).}. 
The filter response curves including the CCD
sensitivity and the atmospheric transmission are shown in Figure 3.

\section{DATA REDUCTION AND SOURCE DETECTION}

\subsection{Data Reduction}

All the individual CCD data were reduced
using IMCAT\footnote{IMCAT is distributed by Nick Keiser at http://www.ifa.hawaii.edu/~kaiser/imcat/}
by the standard process;
bias subtraction, flat fielding, 
combining the frames, astrometry, and photometry.
Here we note that the night sky subtraction needs
two steps to account for fringing and scattered light.
The scattered light also makes it difficult to
carry out accurate flat fielding. However,
our careful experiments show that the central 26$^\prime$
area of the field of view is stable at flat to 1\%
although the scattered light pattern shows variations
as large as a few \% outside the central 26$^\prime$ area.
Details of these reduction procedure are
described in Capak et al. (2006a) and Aussel et al. (2006).
The PSF sizes of final images are summarized in Table 5.
Note that the PSF size of the images used for generating
the official catalogue was matched
to that for the image with worst seeing (1\farcs6).

In Figure 4, we show the composite color image of the whole COSMOS field
made from $B$, $r^\prime$, and $z^\prime$ data.
The reduced images were divided into tiles with a
dimension of $10^{\prime}\times10^{\prime}$ as shown in Figure 5.
Note that the center position of the COSMOS field is located in the \#65 tile.
The region colored in light-blue covered by 81 (9$\times$9) tiles
is the COSMOS HST/ACS field.

We estimate the limiting magnitudes by using the 81 tiles.
For each tile, we performed aperture photometry
for 10000 random points (810000 points in total)
on the PSF matched image (1\farcs6)
with $2^{\prime\prime}$ diameter and $3^{\prime\prime}$ diameter.
Then we evaluated the limiting magnitudes from the standard deviation
for the distribution of the random photometry.
The results are summarized in Table 5.
As shown in Table 5, the 3$\sigma$ limiting magnitudes
are deeper than 27 mag.\ ($2^{\prime\prime}$ aperture)
in $B$, $g^\prime$, $V$, and $r^\prime$.
However, in $z^\prime$ band, the limiting magnitude reaches to 25.7 mag.
Difference of the limiting magnitudes among the 81 tiles is
less than 0.25 mag in each band (Figures~6--12).

Note that another type of limiting magnitudes can be estimated
by using background limited numbers. These results are also shown in 
Figures 13--19 provided by Capak et al. (2006a; see their Table 4).

\subsection{Source Detection and Completeness}

In Figures 20 and 21 we show the number counts of detected objects
against the magnitude measured with $2^{\prime\prime}$ aperture
and $3^{\prime\prime}$ aperture
for each band, respectively.
In this analysis, we use SExtractor version 2.3.2 (Bertin \& Arnouts 1996)
with the detection
criteria of 5-pix connection above the 2$\sigma$ significance
and measured aperture magnitude for the detected objects.
The SExtractor parameter setup file used in our analysis is
given in Table 6.
Apparently, the source detection comes to be incomplete
at a shallower magnitude than the limiting magnitude.
This may be interpreted as a result of flux lost from extended sources.
Note that breaks or drops of the number counts in the bright part are
due to saturated objects.

In order to estimate detection completeness,
we have performed a simulation using the IRAF ARTDATA.
We assume that galaxies have two types of light distributions obeying
the exponential law and the de Vaucouleurs' $r^{1/4}$ law.
For each type of galaxies, we generated 200 model galaxies for each
total magnitude interval (0.2 mag) in 9 tiles (\#026, \#029, \#032, \#062, \#065, \#068,
\#098, \#101, and \#104 in Figure 5).
Their sky positions, half-light radius (0\farcs15 to 0\farcs75),
ellipticities (0.3 to 1.0), and position angles (0$^\circ$ to 360$^\circ$)
are randomly determined.
Then these model galaxies are put into the CCD
data together with Poisson noises. After smoothing model-galaxy images
to match to the seeing size, we try to detect them using SExtractor with
the same procedure as that used before.
The detectability of the model galaxies in each band is shown in
Figures 22--28 as a function of total magnitude
and also summarized in Table 7.

In order to compare the input total magnitude of the model galaxies
used in the completeness analysis
with magnitudes obtained by aperture photometry,
we measured the aperture magnitudes of them by
SExtractor.
Figures 29--32 show the relation between the aperture magnitude
($2^{\prime\prime}$ diameter and $3^{\prime\prime}$ diameter)
and the input total magnitude of model galaxies with the profiles of
the exponential law and de Vaucouleurs' law.
The measured aperture magnitudes are always offset
toward fainter against the input total magnitudes.
These offsets are smaller for $3^{\prime\prime}$ 
aperture magnitudes.

\subsection{Concluding Remarks}

We present deep optical imaging observations made with the Suprime-Cam 
on the Subaru Telescope. Our observations cover the seven filter bands
from $B$ to $z^\prime$. These imaging data allow us to investigate 
photometric properties of $\sim$ 1 million galaxies found in the COSMOS field 
together with the high-resolution ACS $I_{814}$ imaging data.
The major COSMOS datasets including the Subaru images
 and catalogs are publicly available (following calibration 
and validation) through the web site for IPAC/IRSA:\\
{\bf \url{http://irsa.ipac.caltech.edu/data/COSMOS/}}.

\vspace{0.5cm}

The HST COSMOS Treasury program was supported through NASA grant HST-GO-09822. 
We gratefully acknowledge the contributions of the entire COSMOS collaboration
consisting of more than 70 scientists. 
More information on the COSMOS survey is available \\ at
{\bf \url{http://www.astro.caltech.edu/~cosmos}}. It is a pleasure the 
acknowledge the excellent services provided by the NASA IPAC/IRSA 
staff (Anastasia Laity, Anastasia Alexov, Bruce Berriman and John Good) 
in providing online archive and server capabilities for the COSMOS datasets.
The COSMOS Science meeting in May 2005 was supported in part by 
the NSF through grant OISE-0456439.
We would also like to thank the staff at the Subaru Telescope
for their invaluable help. In particular, we would like to thank 
Hisanori Furusawa because his professional help as a support scientist
made our Suprime-Cam observations successful. 
This work was financially supported in part by the Ministry
of Education, Culture, Sports, Science, and Technology (Nos. 10044052 
and 10304013), and by JSPS (15340059 and 17253001). SSS and TN
are JSPS fellows.


\clearpage

\begin{deluxetable}{cccccccc}
\tablenum{1}
\tablecaption{Dithering offset for Pattern A}
\tablewidth{0pt}
\tablehead{
\colhead{ID} & 
\colhead{$\Delta$RA\tablenotemark{a}} & 
\colhead{$\Delta$DEC\tablenotemark{b}} &
\colhead{PA (deg)}  &
\colhead{ID} & 
\colhead{$\Delta$RA\tablenotemark{a}} & 
\colhead{$\Delta$DEC\tablenotemark{b}} &
\colhead{PA (deg)} 
}
\startdata
Pa01 & 45    & 31 & 0 & La01 & 31 & $-$45 & 90 \\
Pa02 & 19    & 32 & 0 & La02 & 32 & $-$19 & 90 \\
Pa03 & $-$7  & 33 & 0 & La03 & 33 & 7 & 90 \\
Pa04 & $-$33 & 31 & 0 & La04 & 31 & 33 & 90 \\
Pa05 & 46    & $-$1 & 0 & La05 & $-$1 & $-$46 & 90 \\
Pa06 & 20    & 0 & 0 & La06 & 0 & $-$20 & 90 \\
Pa07 & $-$6  & 1 & 0 & La07 & 1 & 6 & 90 \\
Pa08 & $-$32 & $-$1 & 0 & La08 & $-$1 & 32 & 90 \\
Pa09 & 45    & $-$33 & 0 & La09 & $-$33 & $-$45 & 90 \\
Pa10 & 19    & $-$32 & 0    & La10 & $-$32 & $-$19 & 90 \\
Pa11 & $-$7  & $-$31 & 0  & La11 & $-$31 & 7 & 90 \\
Pa12 & $-$33 & $-$33 & 0 & La12 & $-$33 & 33 & 90 \\
Pb01 & 32    & 33 & 0 & Lb01 & 33 & $-$32 & 90 \\
Pb02 & 6     & 31 & 0 & Lb02 & 31 & $-$6 & 90 \\
Pb03 & $-$20 & 32 & 0 & Lb03 & 32 & 20 & 90 \\
Pb04 & $-46$ & 33 & 0 & Lb04 & 33 & 46 & 90 \\
Pb05 & 33    & 1 & 0 & Lb05 & 1 & $-$33 & 90 \\
Pb06 & 7     & $-$1 & 0 & Lb06 & $-$1 & $-$7 & 90 \\
Pb07 & $-$19 & 0 & 0 & Lb07 & 0 & 19 & 90 \\
Pb08 & $-$45 & 1 & 0 & Lb08 & 1 & 45 & 90 \\
Pb09 & 32    & $-$31 & 0 & Lb09 & $-$31 & $-$32 & 90 \\
Pb10 & 6     & $-$33 & 0 & Lb10 & $-$33 & $-$6 & 90 \\
Pb11 & $-$20 & $-$32 & 0 & Lb11 & $-$32 & 20 & 90 \\
Pb12 & $-$46 & $-$31 & 0 & Lb12 & $-$31 & 46 & 90 \\
\enddata
\tablenotetext{a}{Offset in RA from the COSMOS center position
                  in units of arcmin.}
\tablenotetext{b}{Offset in DEC from the COSMOS center position
                  in units of arcmin.}
\end{deluxetable}

\clearpage
\begin{deluxetable}{cccccccc}
\tablenum{2}
\tablecaption{Dithering offset for Pattern C}
\tablewidth{0pt}
\tablehead{
\colhead{ID} & 
\colhead{$\Delta$RA\tablenotemark{a}} & 
\colhead{$\Delta$DEC\tablenotemark{b}} &
\colhead{PA (deg)} &
\colhead{ID} & 
\colhead{$\Delta$RA\tablenotemark{a}} & 
\colhead{$\Delta$DEC\tablenotemark{b}} &
\colhead{PA (deg)} 
}
\startdata
A1 & $-$3.5 & 26.5 & 0     & C1 & 3.5 & $-$26.5 & 0 \\
A2 & $-$30.5 & 27.5 & 0    & C2 & 30.5 & $-$27.5 & 0 \\
A3 & 30.0 & 0.5 & 0        & C3 & $-$30.0 & $-$0.5 & 0 \\
A4 & 3.0 & $-$0.5 & 0      & C4 & $-$3.0 & 0.5 & 0 \\
A5 & $-$3.5 & $-$27.5 & 0  & C5 & 3.5 & 27.5 & 0 \\
A6 & $-$30.5 & $-$26.5 & 0 & C6 & 30.5 & 26.5 & 0 \\
A7 & 26.0  & 30.0 & 90     & C7 & $-$26.0 & $-$30.0 & 90 \\
A8 & $-$27.0 & 3.0 & 90    & C8 & 27.0 & $-$3.0 & 90 \\
A9 & 27.0  & $-$30.0 & 90  & C9 & $-$27.0 & 30.0 & 90 \\
B1 & 26.5  & 3.5 & 90      & D1 & $-$26.5 & $-$3.5 & 90 \\
B2 & 27.5  & 30.5 & 90     & D2 & $-$27.5 & $-$30.5 & 90 \\
B3 & 0.5 & $-$30.0 & 90    & D3 & $-$0.5 & 30.0 & 90 \\
B4 & $-$0.5 & $-$3.0 & 90  & D4 & 0.5 & 3.0 & 90 \\
B5 & $-$27.5 & 3.5 & 90    & D5 & 27.5 & $-$3.5 & 90 \\
B6 & $-$26.5 & 30.5 & 90  & D6 & 26.5 & $-$30.5 & 90 \\
B7 & 30.0 & $-$26.0 & 0    & D7 & $-$30.0 & 26.0 & 0 \\
B8 & 3.0 & 27.0 & 0        & D8 & $-$3.0 & $-$27.0 & 0 \\
B9 & $-$30.0 & $-$27.0 & 0 & D9 & 30.0 & 27.0 & 0 \\
\enddata
\tablenotetext{a}{Offset in RA from the COSMOS center position
                  in units of arcmin.}
\tablenotetext{b}{Offset in DEC from the COSMOS center position
                  in units of arcmin.}
\end{deluxetable}

\begin{deluxetable}{ccccc}
\tablenum{3}
\tablecaption{A summary of observational programs}
\tablewidth{0pt}
\tablehead{
\colhead{Semester} & 
\colhead{ID No.} & 
\colhead{PI}  &
\colhead{Program Title} &
\colhead{Nights} 
}
\startdata
S03B & 239 & Y. Taniguchi & COSMOS-Broad\tablenotemark{a}  & 10  \\
S04A & 080 & Y. Taniguchi & COSMOS-Narrow\tablenotemark{b} & 2.5 \\
S04B & 142 & Y. Taniguchi & COSMOS-21\tablenotemark{c}     & 8   \\
S04B & UH-17A & N. Scoville & COSMOS-21\tablenotemark{c}   & 4   \\
\enddata
\tablenotetext{a}{Suprime-Cam Imaging of the HST COSMOS 2-Degree ACS 
                  Survey Deep Field (Intensive Program).}
\tablenotetext{b}{Wide-Field Search for Ly$\alpha$ Emitters at $z$=5.7 
                  in the HST/COSMOS Field.}
\tablenotetext{c}{COSMOS-21: Deep Intermediate \& Narrow-band Survey of 
                  the COSMOS Field (Intensive Program). This proposal was
                  also granted in S05B (S05B-013). However, this COSMOS-21
                  program will be described in a forthcoming paper.}
\end{deluxetable}

\begin{deluxetable}{ccccc}
\tablenum{4}
\tablecaption{A summary of observational runs}
\tablewidth{0pt}
\tablehead{
\colhead{ID No.} & 
\colhead{Period} & 
\colhead{Nights}  &
\colhead{Avail. Nights} &
\colhead{Bands} 
}
\startdata
S03B-239 & 2004 Jan 16-21 & 6   & 5 & $B$, $r^\prime$, $i^\prime$, $z^\prime$ \\
S03B-239 & 2004 Feb 15-18 & 4   & 2 & $V$, $i^\prime$ \\
S04A-080 & 2004 Apr 15-19\tablenotemark{a} & 2.5 & 1 & $NB816$ \\
S04B-142 & 2005 Jan 8-10  & 3   & 0 & no obs. \\
UH-17A   & 2005 Feb 3     & 1   & 0 & no obs. \\
S04B-142 & 2005 Feb 9-13  & 5   & 2 & $g^\prime$, $V$, $NB816$ \\
UH-17A   & 2005 Mar 10-12 & 3   & 1 & $NB816$ \\
S04B-142 & 2005 Apr 1-4\tablenotemark{b} & 4 & 3 & $g^\prime$, $NB816$ \\
\enddata
\tablenotetext{a}{First half night was used in every night.}
\tablenotetext{b}{Compensation nights because of the poor weather in
         S04B-142 Jan and Feb runs.}
\end{deluxetable}

\begin{deluxetable}{ccccccc}
\tablenum{5}
\tablecaption{A summary of the optical imaging data for COSMOS.}
\tablewidth{0pt}
\tablehead{
\colhead{Band} & 
\colhead{$\lambda_{\rm c}$\tablenotemark{a}} & 
\colhead{$\Delta\lambda$\tablenotemark{b}} & 
\colhead{Total TDT\tablenotemark{c}} & 
\colhead{$m_{\rm lim}$\tablenotemark{d}}  &
\colhead{$m_{\rm lim}$\tablenotemark{e}}   &
\colhead{$FWHM({\rm PSF})$\tablenotemark{f}}   \\
\colhead{} & 
\colhead{(\AA)} &
\colhead{(\AA)} &
\colhead{(min)} & 
\colhead{(mag)}  &
\colhead{(mag)}  &
\colhead{($^{\prime\prime}$)} 
}
\startdata
$B$        & 4459.7 & 897  & 70.3  & 27.8   & 27.2 & 0.95 \\
$g^\prime$ & 4479.6 & 1265 & 86.0  & 27.2   & 26.6 & 1.58 \\
$V$        & 5483.8 & 946  & 50.3  & 27.1   & 26.5 & 1.33 \\
$r^\prime$ & 6295.1 & 1382 & 36.0  & 27.2   & 26.6 & 1.05 \\
$i^\prime$ & 7640.8 & 1497 & 40.3  & 26.8   & 26.1 & 0.95 \\
$z^\prime$ & 9036.9 & 856  & 63.5  & 25.9   & 25.3 & 1.15 \\
$NB816$    & 8151.0 & 117  & 187.7 & 26.1   & 25.7 & 1.51 \\
\enddata
\tablenotetext{a}{Central wavelength.}
\tablenotetext{b}{Filter band width.}
\tablenotetext{c}{The total target dedicated time.}
\tablenotetext{d}{The 3$\sigma$ limiting magnitude in the AB system within
		$2^{\prime\prime}$ diameter aperture.}
\tablenotetext{e}{The 3$\sigma$ limiting magnitude in the AB system within
		$3^{\prime\prime}$ diameter aperture.}
\tablenotetext{f}{The PSF size of the final images. Note that
              the PSF size of each filter band is finally matched into 1\farcs6 
              in the official photometric catalogue.}

\end{deluxetable}


\begin{deluxetable}{lll}
\tablenum{6}
\tablecaption{SExtractor Parameters \label{t:sex-param}}
\tablehead{
\colhead{Parameter} & \colhead{Setting} & \colhead{Comment}
}
\startdata
PARAMETERS\_NAME & cosmos-subaru.param & Fields to be included in output catalog\\
FILTER\_NAME & gauss\_2.5\_5x5.conv & Filter for detection image\\
STARNNW\_NAME & default.nnw & Neural-Network\_Weight table filename \\
CATALOG\_NAME & STDOUT & Output to pipe instead of file \\
CATALOG\_TYPE & ASCII & Output type\\
DETECT\_TYPE & CCD & Detector type\\
DETECT\_MINAREA & 5 & Minimum number of pixels above threshold\\
DETECT\_THRESH & 2 & Detection Threshold in $\sigma$\\
ANALYSIS\_THRESH & 2 & Limit for isophotal analysis $\sigma$\\
FILTER & Y & Use filtering \\
DEBLEND\_NTHRESH & 64 & Number of deblending sub-thresholds\\
DEBLEND\_MINCONT & 0.0 & Minimum contrast parameter for deblending\\
CLEAN & Y & Clean spurious detections\\
CLEAN\_PARAM & 1 & Cleaning efficiency\\
MASK\_TYPE & CORRECT & Correct flux for blended objects\\
PHOT\_APERTURES & 13.3, 20 & MAG\_APER aperture diameter(s) in pixels\\
PHOT\_AUTOPARAMS & 2.5, 3.5 & MAG\_AUTO parameters: $<$Kron\_fact$>$,$<$min\_radius$>$\\
PHOT\_FLUXFRAC & 0.2,0.5,0.8 & Define n-light radii\\
PHOT\_AUTOAPERS & 20.0, 20.0 & MAG\_AUTO minimum apertures: estimation, photometry\\
SATUR\_LEVEL & 300000 & Level of saturation\\
MAG\_ZEROPOINT & 31.4 & Magnitude zero-point\\
GAIN & 1 & Gain is 1 for absolute RMS map\\
PIXEL\_SCALE & 0 & Size of pixel in $^{\prime\prime}$\\
SEEING\_FWHM & 1.5 & Stellar FWHM in $^{\prime\prime}$\\
BACK\_SIZE & 256 & Background mesh in pixels \\
BACK\_FILTERSIZE & 5 & Background filter\\
BACKPHOTO\_TYPE & GLOBAL & Photometry background subtraction type\\
BACKPHOTO\_THICK & 8 & Thickness of the background LOCAL annulus\\
WEIGHT\_GAIN & N & Gain does not vary with changes in RMS noise\\
WEIGHT\_TYPE & MAP\_RMS & Set Weight image type\\
MEMORY\_PIXSTACK & 1000000 & Number of pixels in stack\\
MEMORY\_BUFSIZE & 4096 & Number of lines in buffer\\
MEMORY\_OBJSTACK & 60000 & Size of the buffer containing objects\\
VERBOSE\_TYPE & QUIET & \\
\enddata
\end{deluxetable}

\begin{deluxetable}{ccccccc}
\tablenum{7}
\tablecaption{Results of analysis for the detection completeness.}
\tablewidth{0pt}
\tablehead{
\colhead{Band} & 
\multicolumn{3}{c}{Exponential Law\tablenotemark{a}} &
\multicolumn{3}{c}{de Vaucouleurs'  Law\tablenotemark{b}} \\
\colhead{} & \colhead{95\%} & \colhead{90\%} & \colhead{50\%} &
   \colhead{95\%} & \colhead{90\%} & \colhead{50\%} \\
\colhead{} & \colhead{(mag)} & \colhead{(mag)}  & \colhead{(mag)}  &
\colhead{(mag)} & \colhead{(mag)} & \colhead{(mag)} }
\startdata
$B$        & 24.1 & 24.7 & 25.5 & 24.1 & 24.7 & 25.3 \\
$g^\prime$ & 23.9 & 24.3 & 24.9 & 23.7 & 24.3 & 24.7 \\
$V$        & 23.5 & 24.1 & 24.7 & 23.3 & 24.1 & 24.7 \\
$r^\prime$ & 23.5 & 24.1 & 24.7 & 23.3 & 23.7 & 24.7 \\
$i^\prime$ & 22.5 & 23.5 & 24.3 & 22.7 & 23.3 & 24.1 \\
$z^\prime$ & 22.1 & 22.9 & 23.3 & 22.1 & 22.7 & 23.1 \\
$NB816$    & 22.1 & 23.1 & 23.7 & 22.1 & 22.9 & 23.5 \\
\enddata
\tablenotetext{a}{Magnitude at which the detection completeness is greater than
   95\%, 90\%, and 50\%
           for the model galaxies with the exponential light profile.}
\tablenotetext{b}{Magnitude at which the detection completeness is greater than
   95\%, 90\%, and 50\%
           for the model galaxies with the de Vaucouleurs' law light profile.}
\end{deluxetable}

\clearpage

\begin{figure}[ht]
\plotone{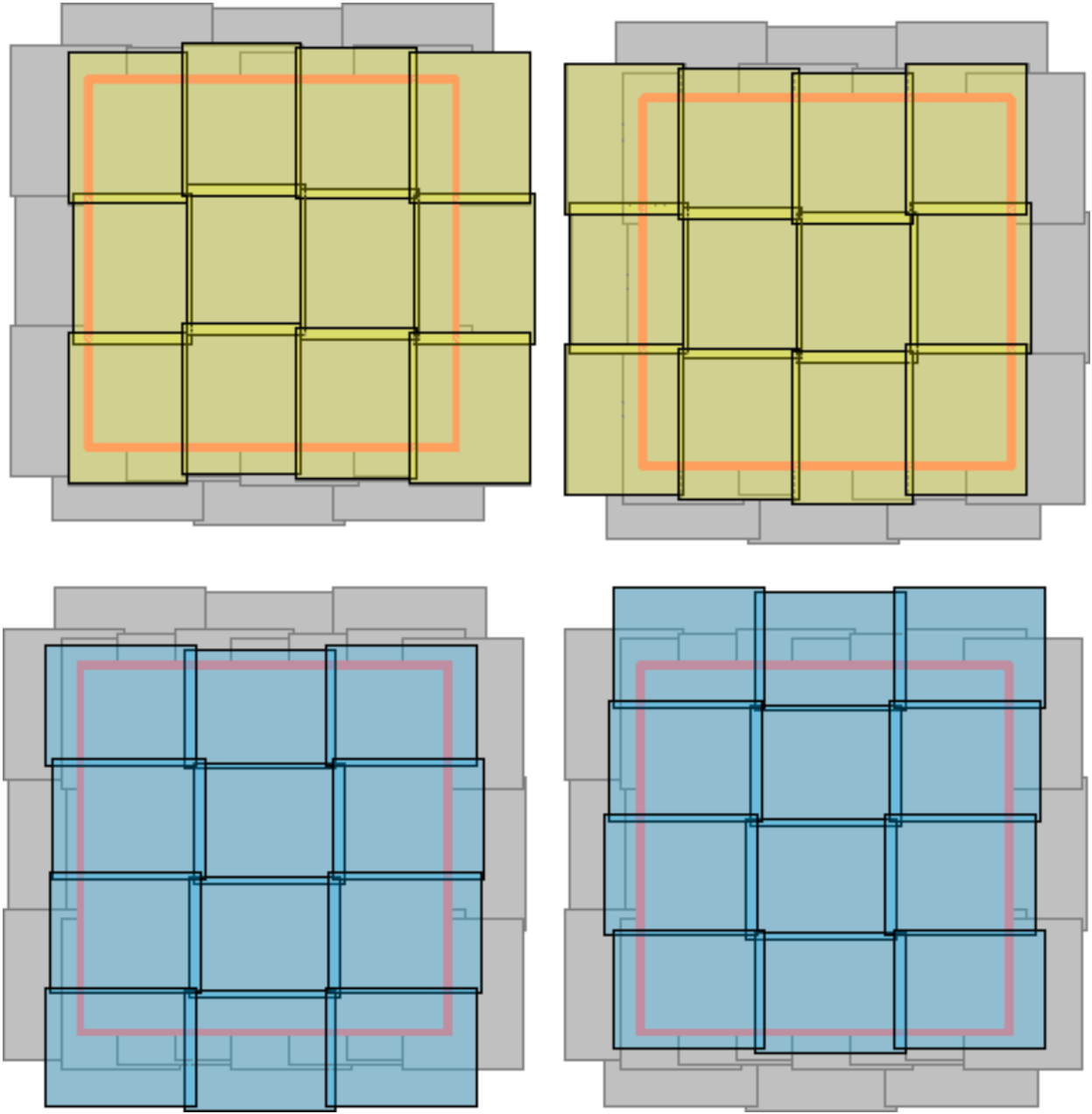}
\caption{Dithering Pattern A.
\label{DithA}}
\end{figure}
\clearpage

\begin{figure}[ht]
\plotone{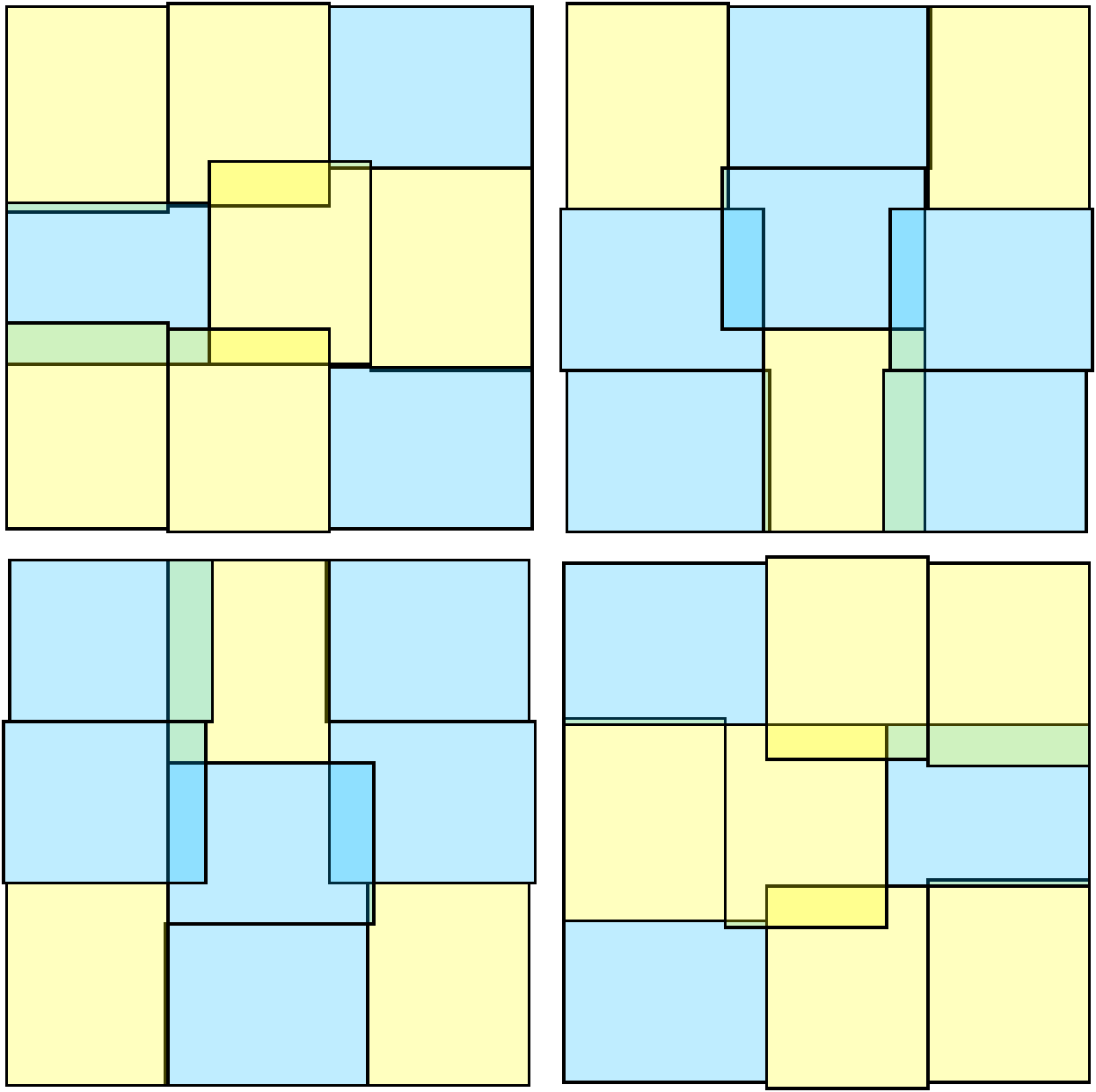}
\caption{Dithering Pattern C.
\label{DithC}}
\end{figure}
\clearpage

\begin{figure}[ht]
\plotone{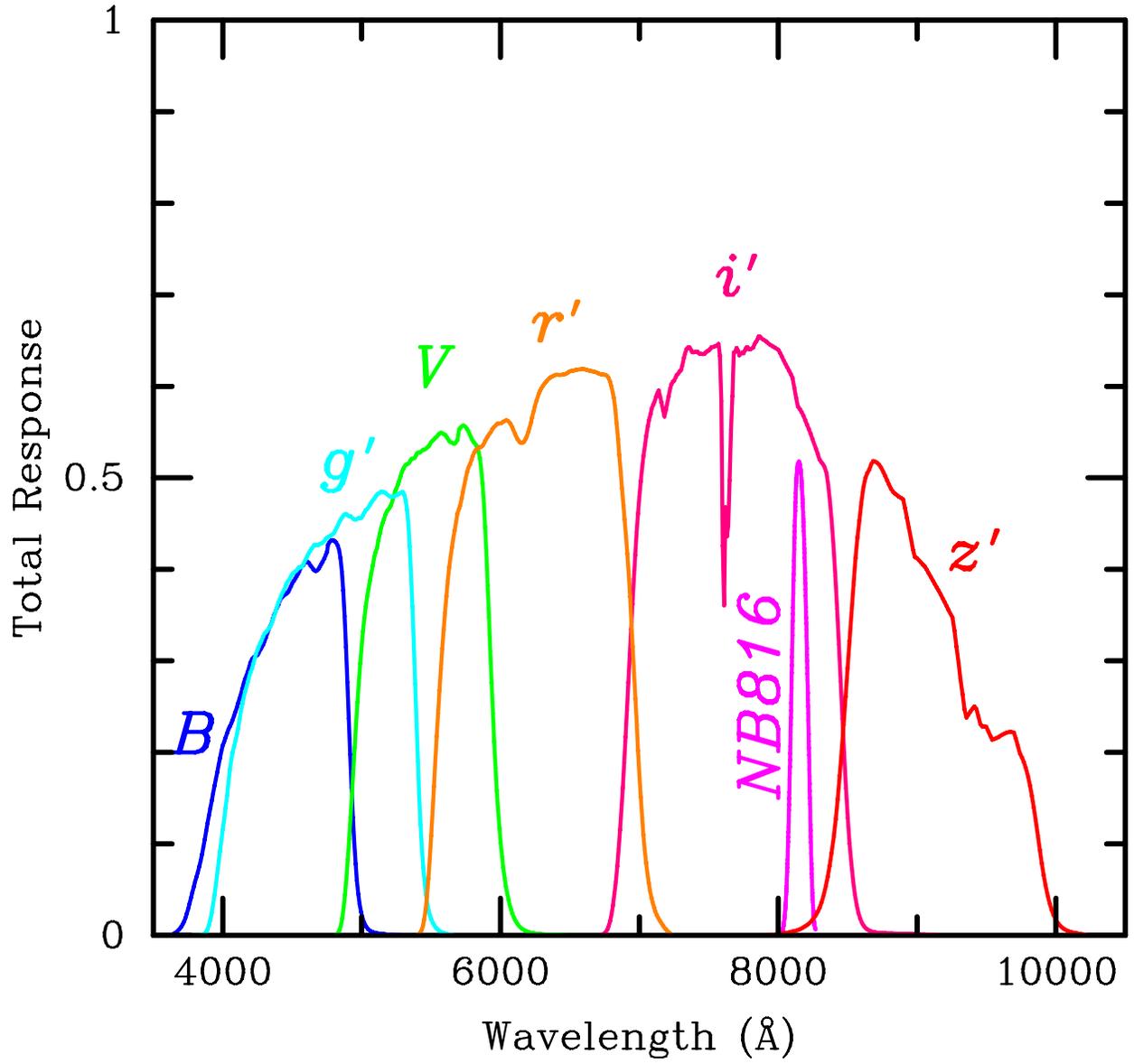}
\caption{Filter response curves including the CCD
sensitivity and the atmospheric transmission.
\label{Resp}}
\end{figure}
\clearpage

\begin{figure}[ht]
\plotone{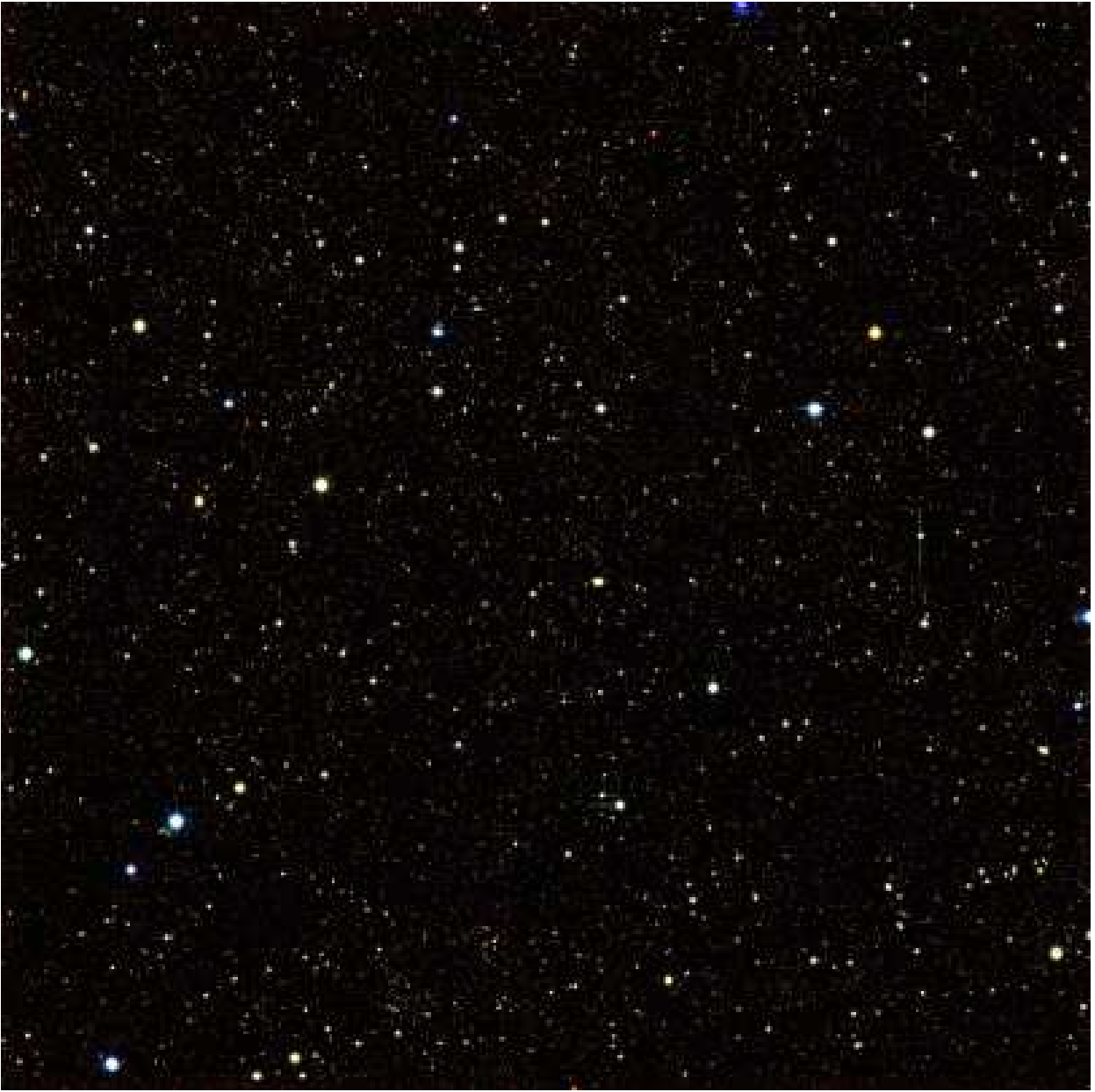}
\caption{A color image of the COSMOS field made from
$B$, $r^\prime$, and $z^\prime$ data.
The image size is 1.5$^\circ$ $\times$ 1.5$^\circ$.
\label{Color}}
\end{figure}
\clearpage

\begin{figure}[ht]
\plotone{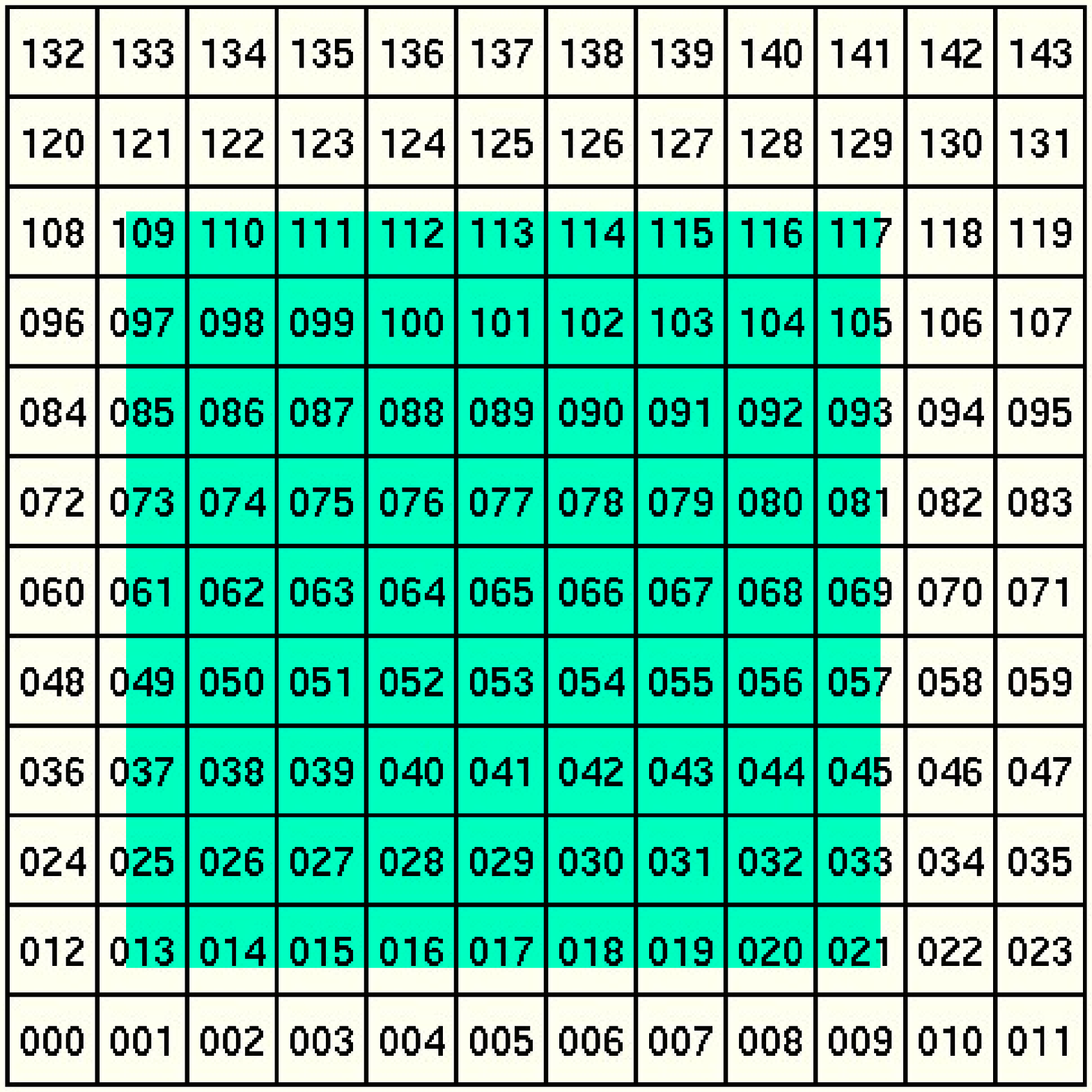}
\caption{Sub-tiles of the reduced images.
Each tile has a $10^\prime \times 10^\prime$ dimension.
The region with light-blue color is the COSMOS 2 square degree field
covered by our HST/ACS observations.
The field center position is located at \#65 tile.
\label{Subtile}}
\end{figure}
\clearpage

\begin{figure}[ht]
\plotone{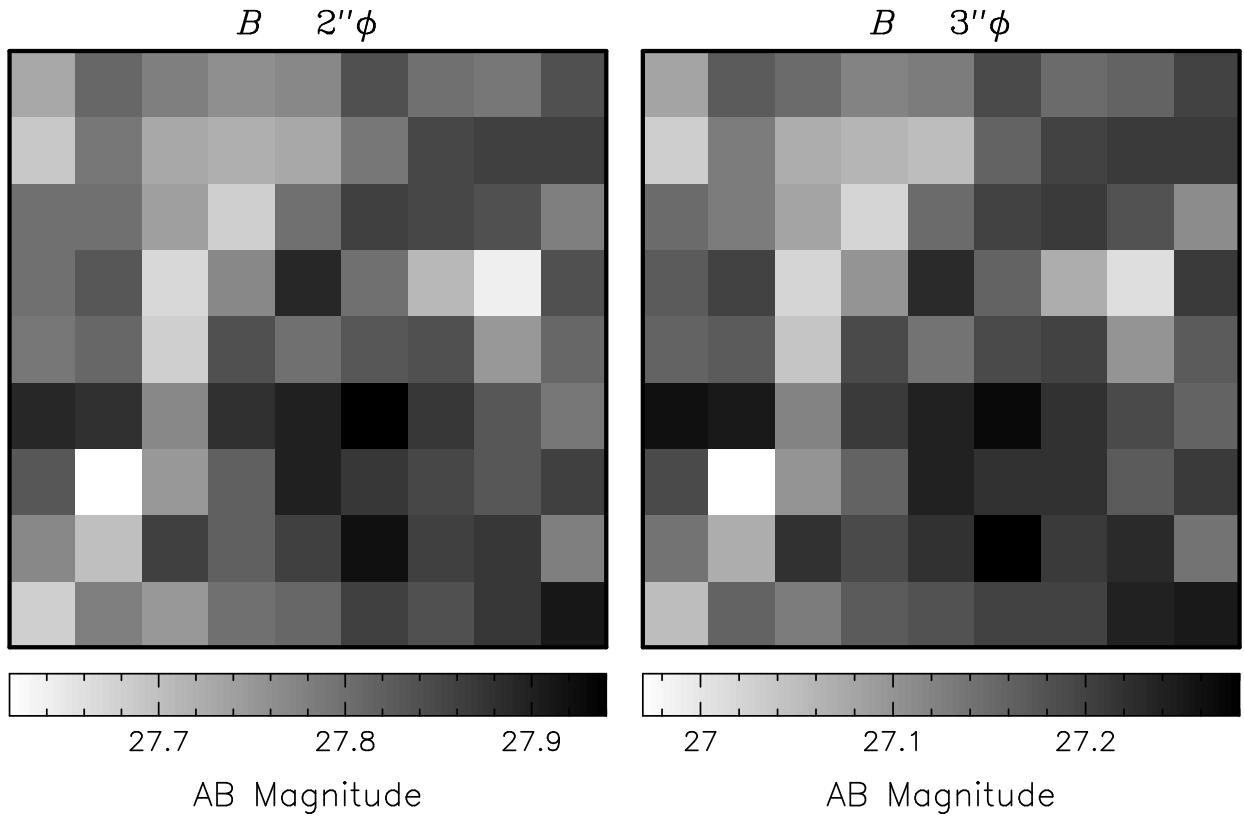}
\caption{Variation of 3$\sigma$ limiting magnitudes
in the $B$ band among the 81 tiles covering the COSMOS HST/ACS region
(as shown by the region with light-blue color in Figure \ref{Subtile}).
The image size is 1.5$^\circ$ $\times$ 1.5$^\circ$.
The left panel shows the case for  $2^{\prime\prime}$ diameter aperture and
the right panel is  for  $3^{\prime\prime}$ diameter aperture.
\label{limb}}
\end{figure}
\clearpage

\begin{figure}[ht]
\plotone{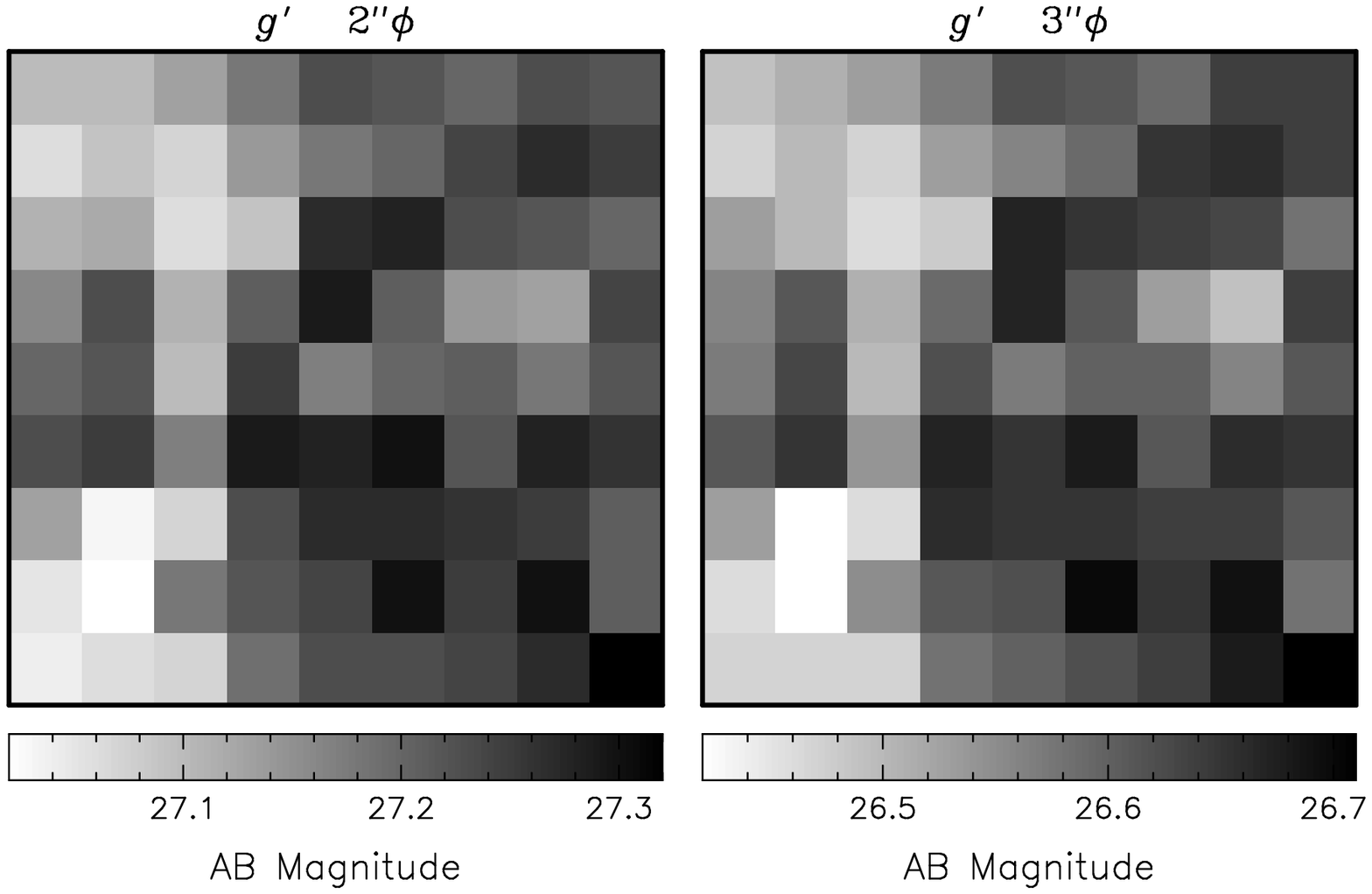}
\caption{Same as Figure \ref{limb} for the $g^{\prime}$ band.
\label{limgp}}
\end{figure}
\clearpage

\begin{figure}[ht]
\plotone{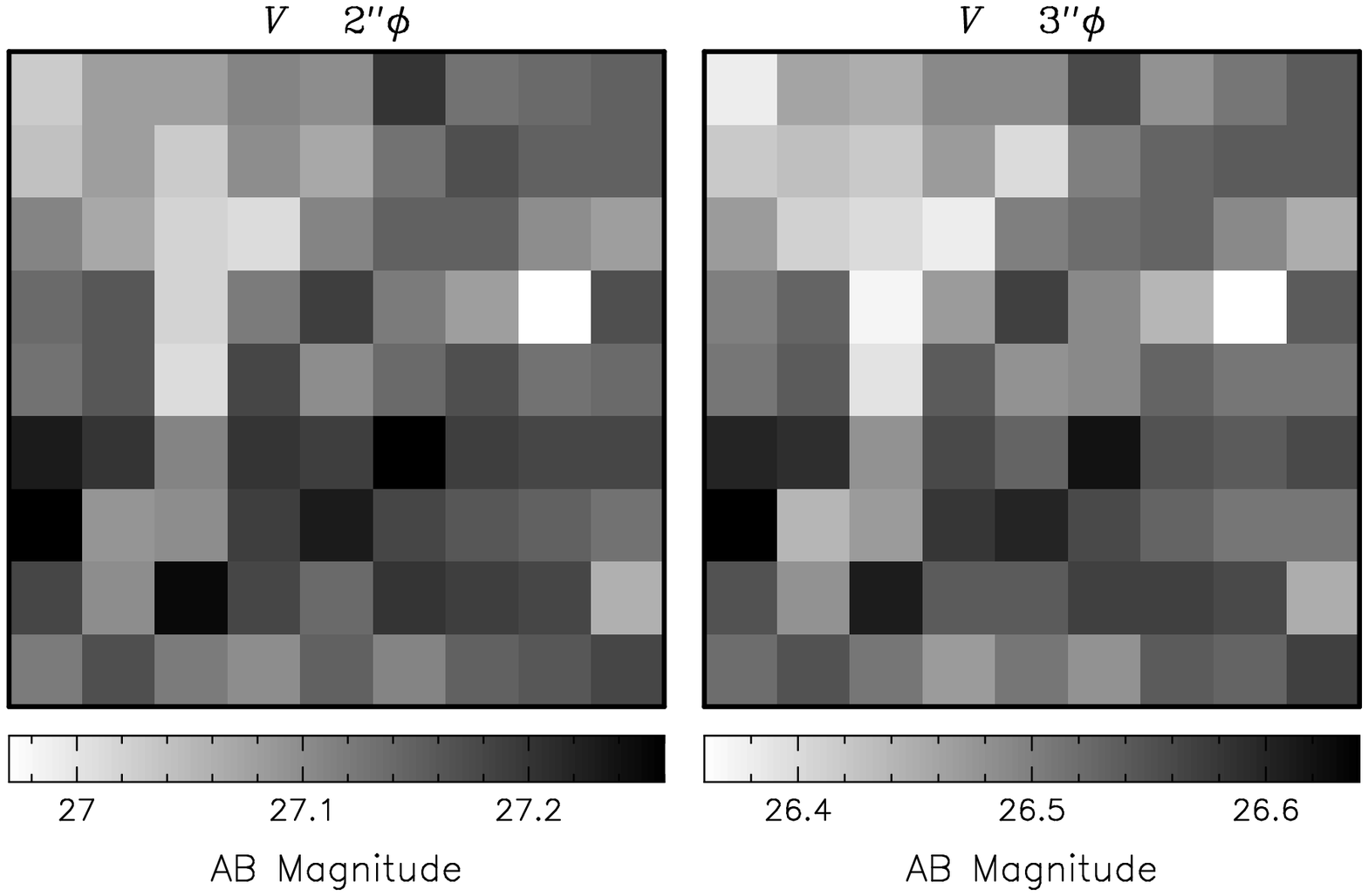}
\caption{Same as Figure \ref{limb} for the $V$ band.
\label{limv}}
\end{figure}
\clearpage

\begin{figure}[ht]
\plotone{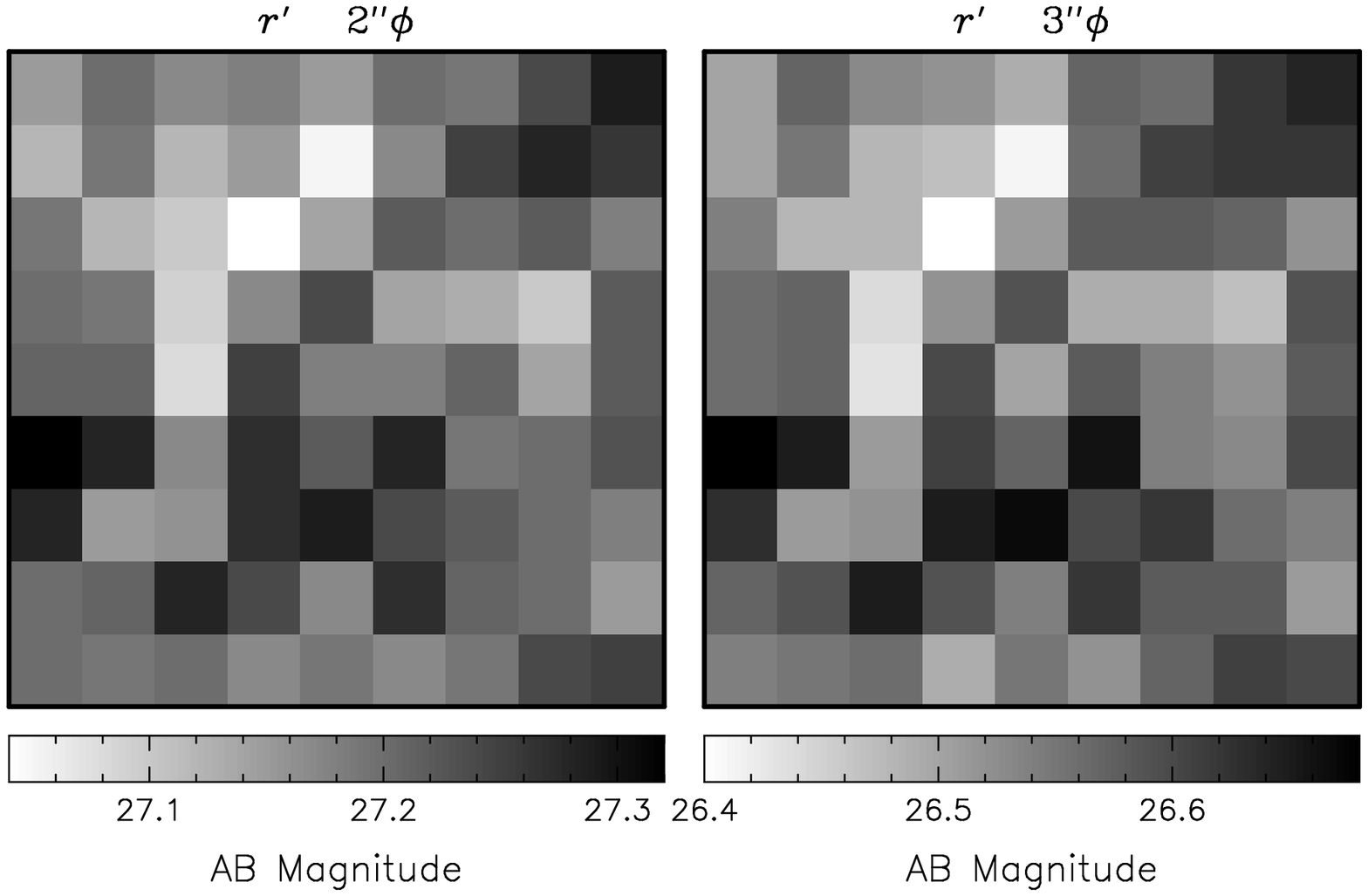}
\caption{Same as Figure \ref{limb} for the $r^{\prime}$ band.
\label{limrp}}
\end{figure}
\clearpage

\begin{figure}[ht]
\plotone{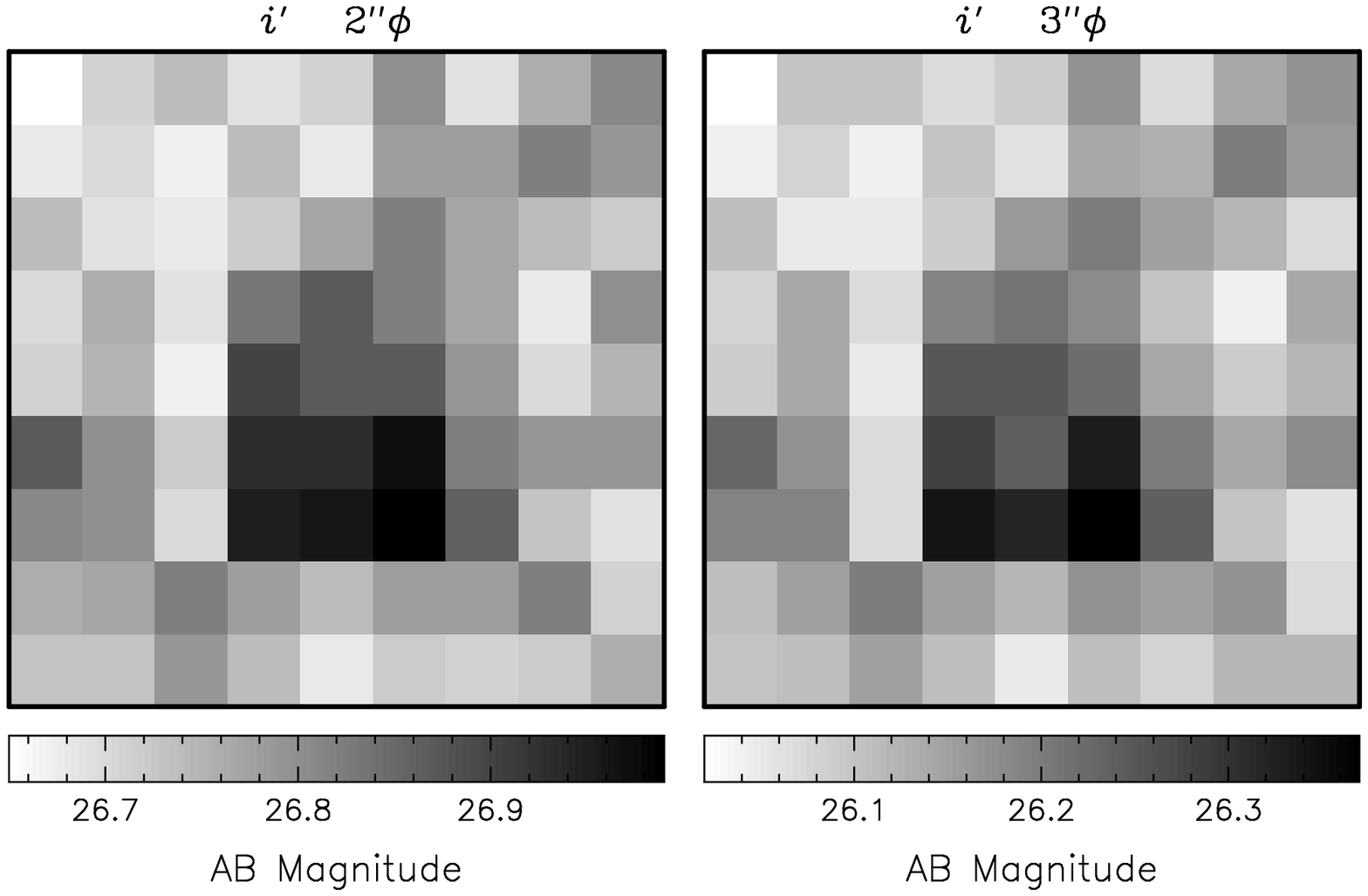}
\caption{Same as Figure \ref{limb} for the $i^{\prime}$ band.
\label{limip}}
\end{figure}
\clearpage

\begin{figure}[ht]
\plotone{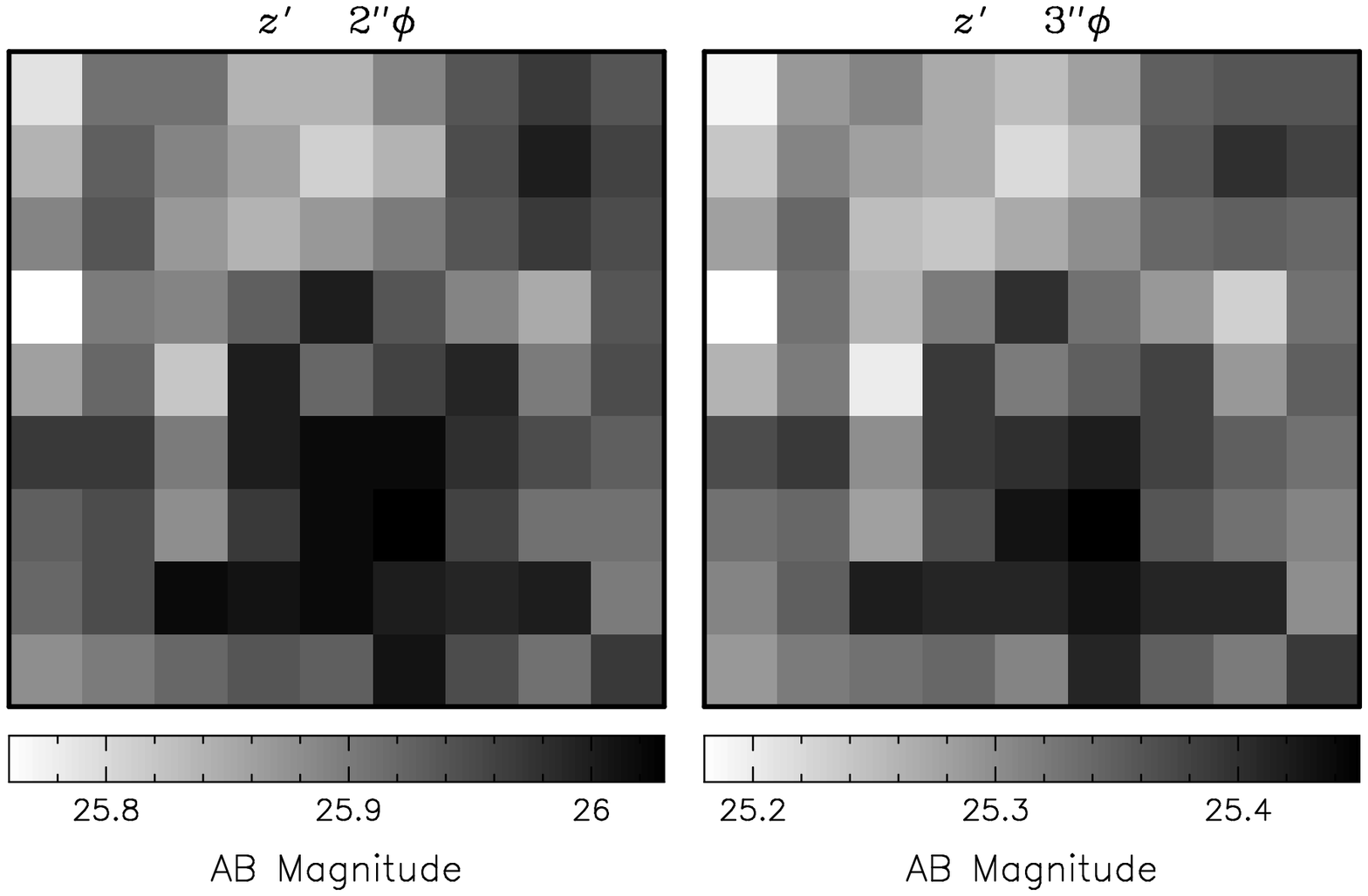}
\caption{Same as Figure \ref{limb} for the $z^{\prime}$ band.
\label{limzp}}
\end{figure}
\clearpage

\begin{figure}[ht]
\plotone{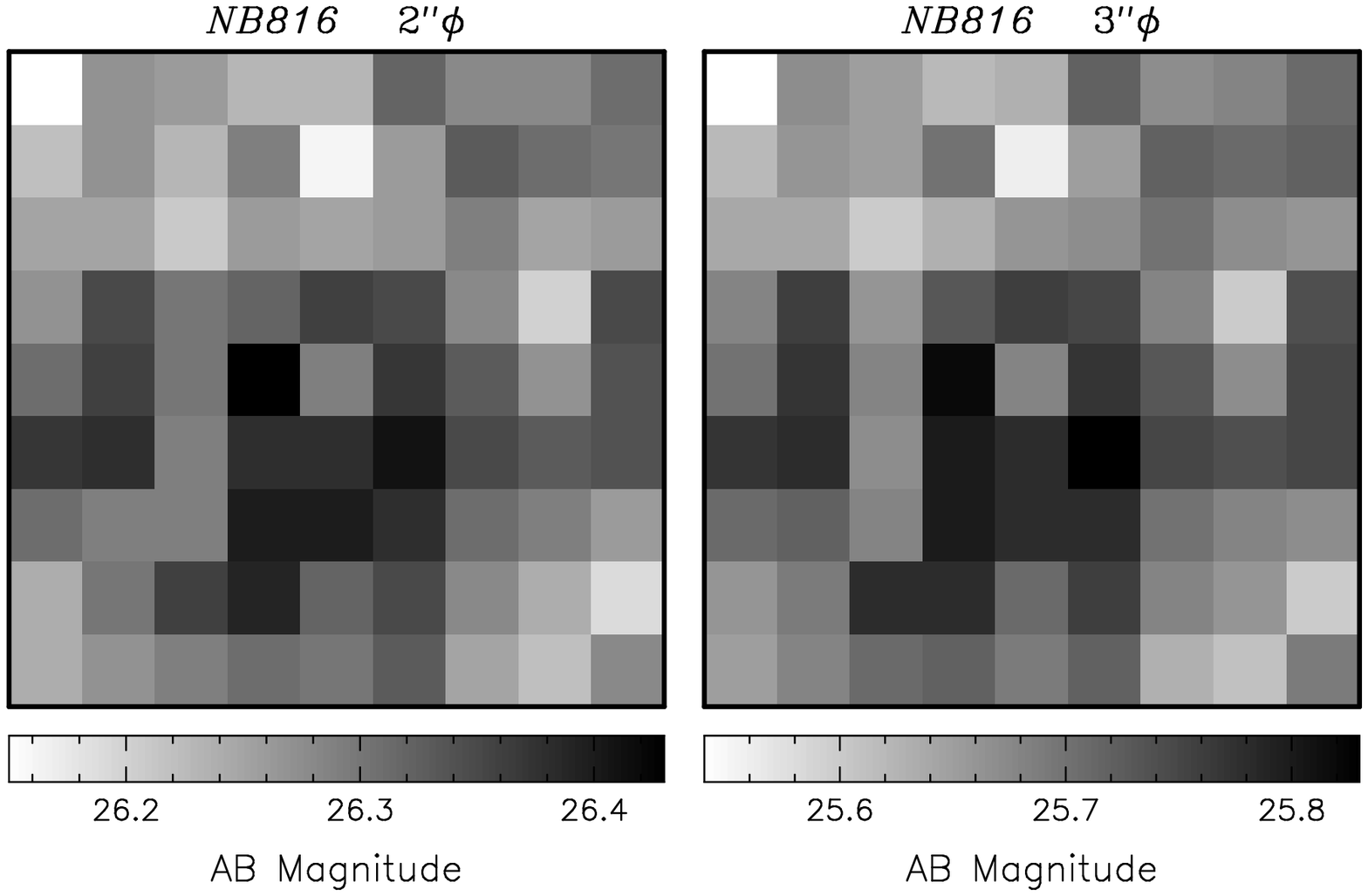}
\caption{Same as Figure \ref{limb} for the $NB816$ band.
\label{limnb}}
\end{figure}
\clearpage

\begin{figure}[ht]
\epsscale{0.45}
\plotone{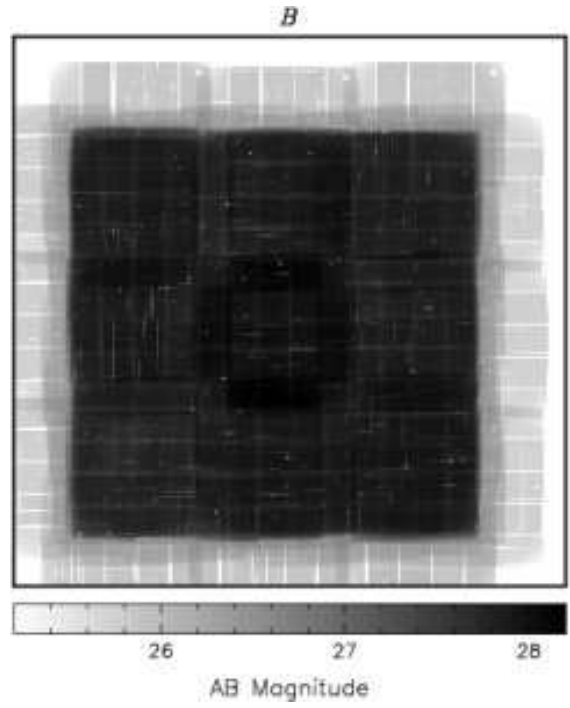}
\caption{The map of 3$\sigma$ limiting magnitude for $3^{\prime\prime}$ diameter aperture
in the $B$ band estimated by Capak et al.~(2006a).
The image size is 2$^\circ$ $\times$ 2$^\circ$.
\label{DepthB}}
\end{figure}
\clearpage

\begin{figure}[ht]
\epsscale{0.45}
\plotone{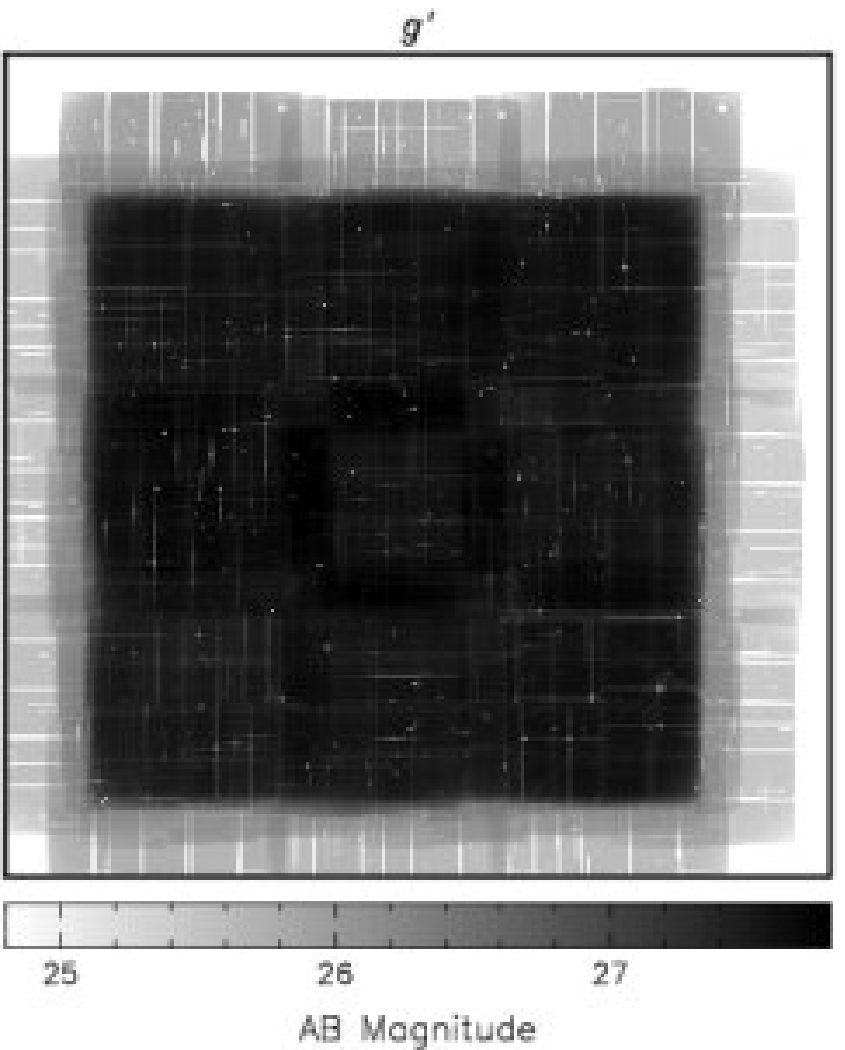}
\caption{Same as Figure \ref{DepthB} for  the $g^{\prime}$ band.
\label{Depthgp}}
\end{figure}
\clearpage

\begin{figure}[ht]
\epsscale{0.45}
\plotone{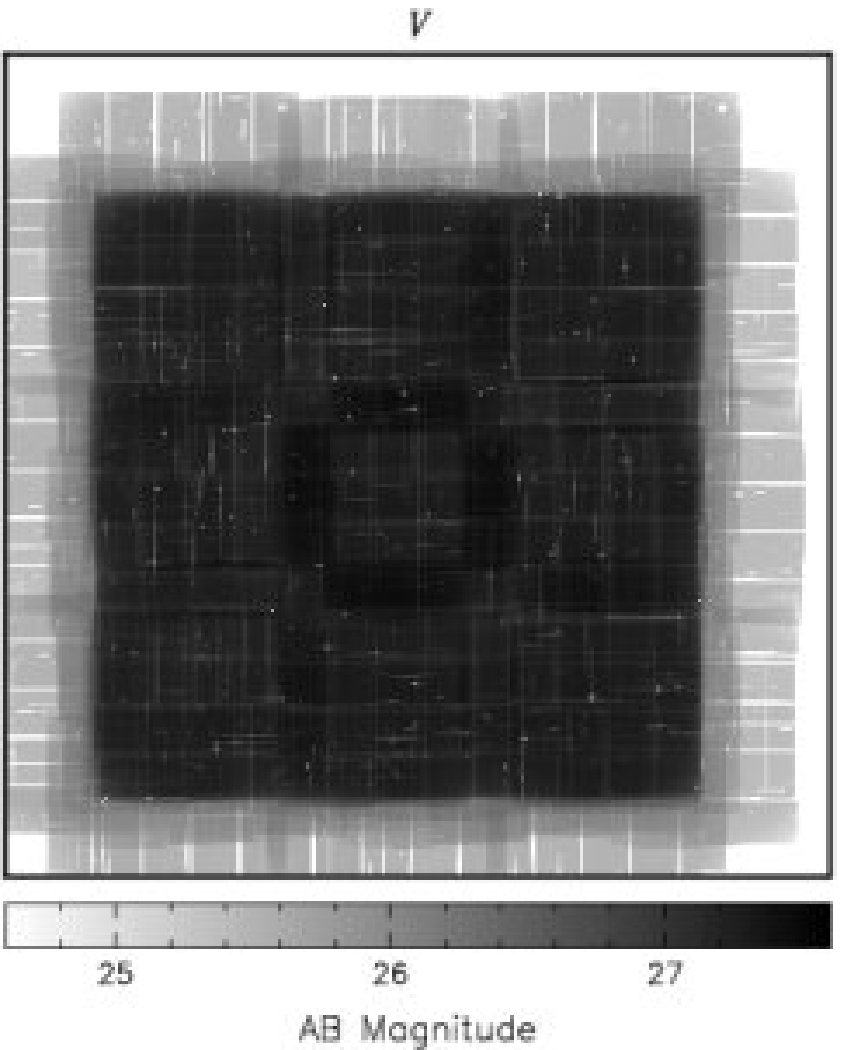}
\caption{Same as Figure \ref{DepthB} for  the $V$ band.
\label{Depthv}}
\end{figure}
\clearpage

\begin{figure}[ht]
\epsscale{0.45}
\plotone{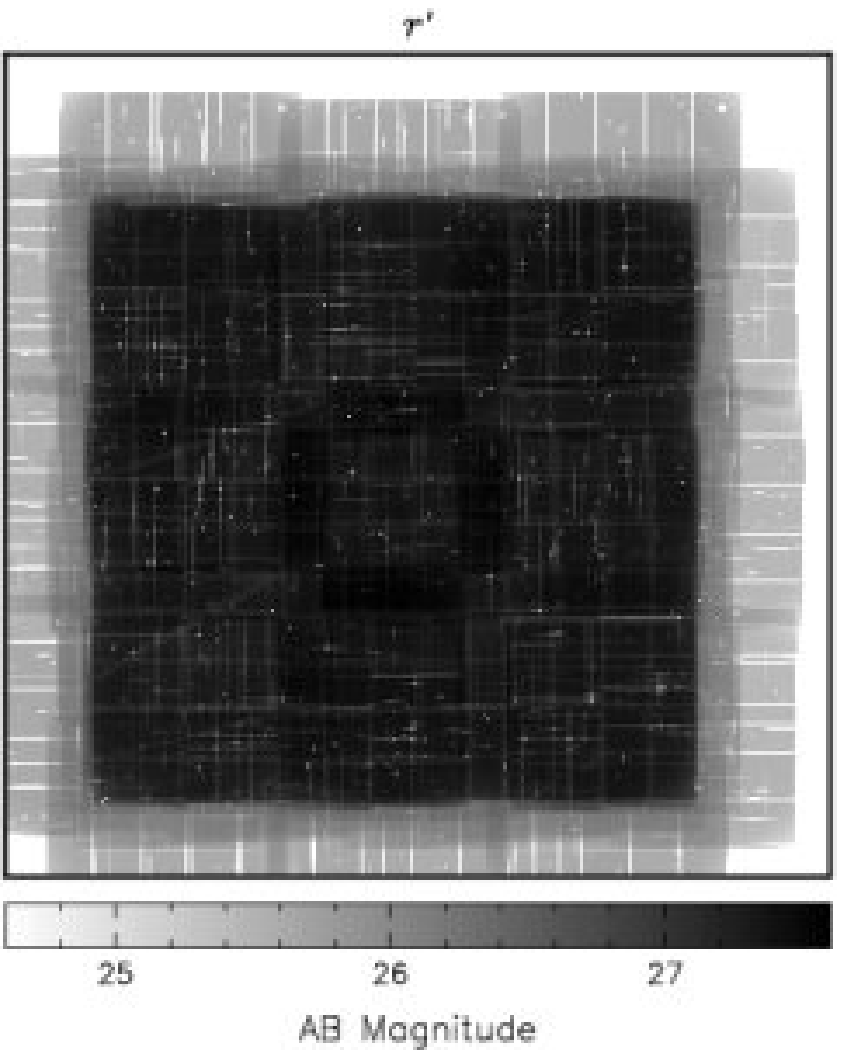}
\caption{Same as Figure \ref{DepthB} for  the $r^{\prime}$ band.
\label{Depthrp}}
\end{figure}
\clearpage

\begin{figure}[ht]
\epsscale{0.45}
\plotone{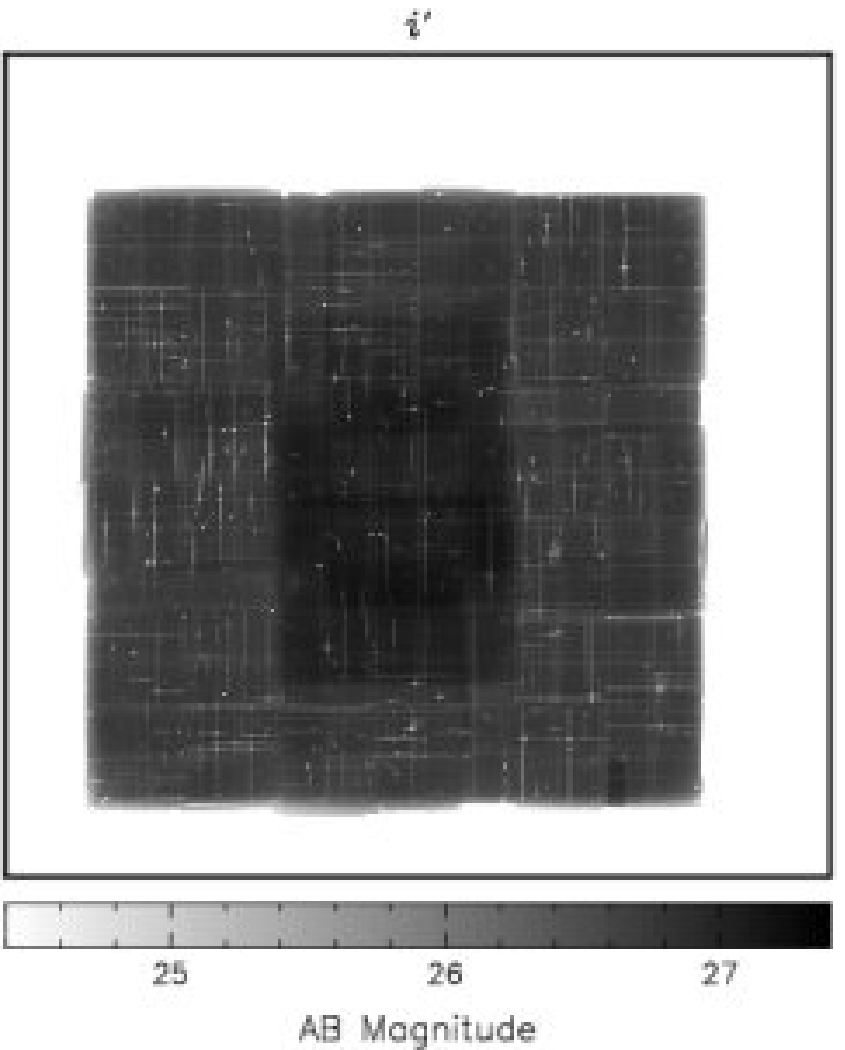}
\caption{Same as Figure \ref{DepthB} for  the $i^{\prime}$ band.
\label{Depthip}}
\end{figure}
\clearpage

\begin{figure}[ht]
\epsscale{0.45}
\plotone{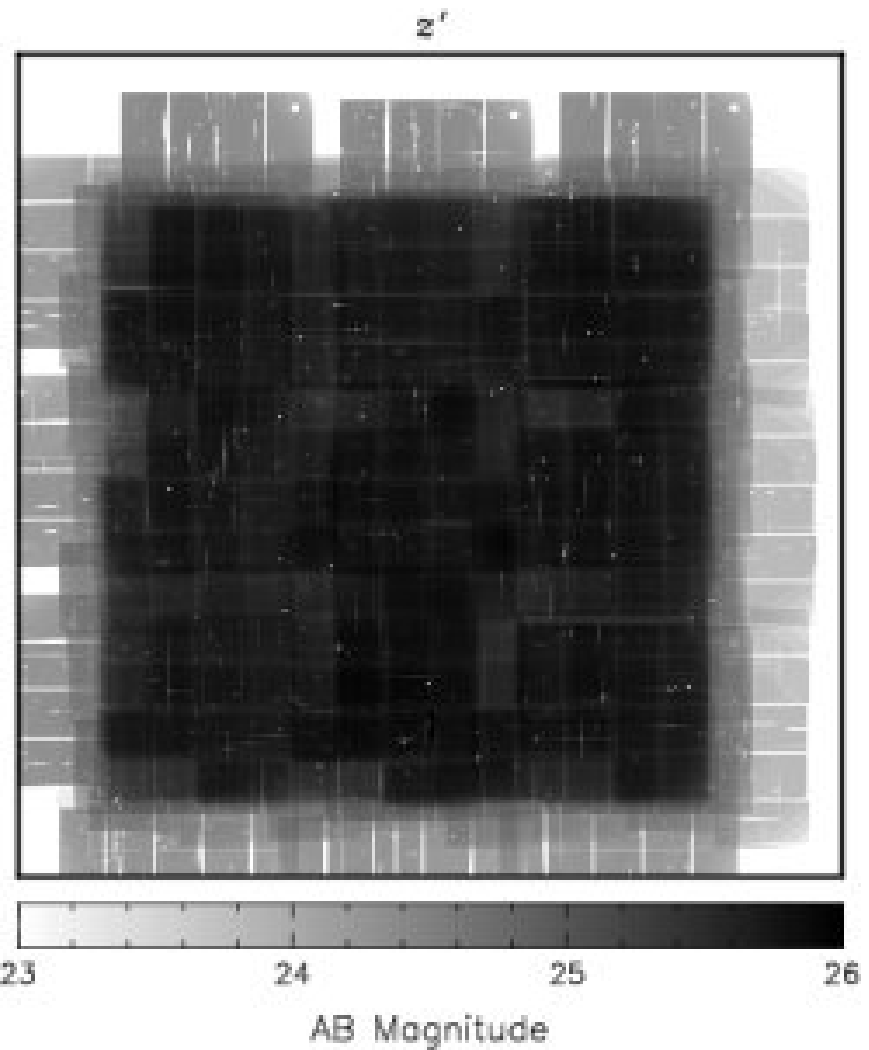}
\caption{Same as Figure \ref{DepthB} for  the $z^{\prime}$ band.
\label{Depthzp}}
\end{figure}
\clearpage

\begin{figure}[ht]
\epsscale{0.45}
\plotone{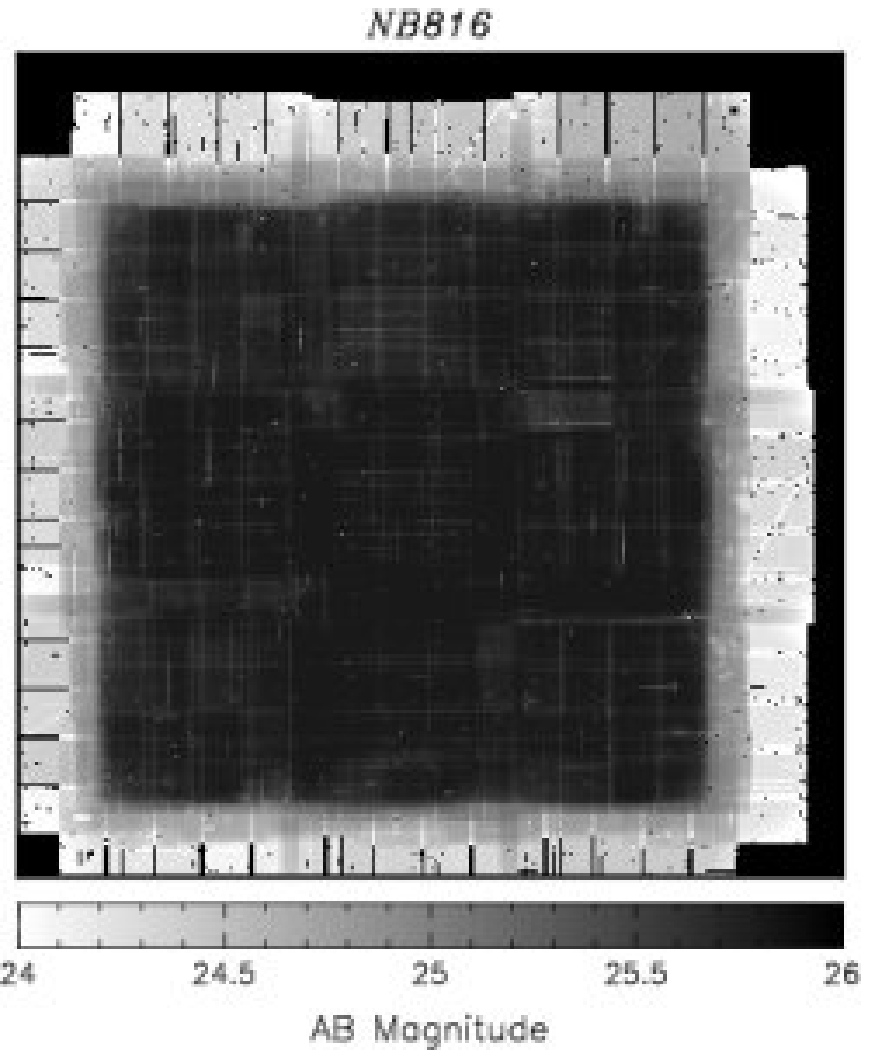}
\caption{Same as Figure \ref{DepthB} for  the $NB816$ band.
\label{Depthnb}}
\end{figure}
\clearpage

\begin{figure}[ht]
\epsscale{0.6}
\plotone{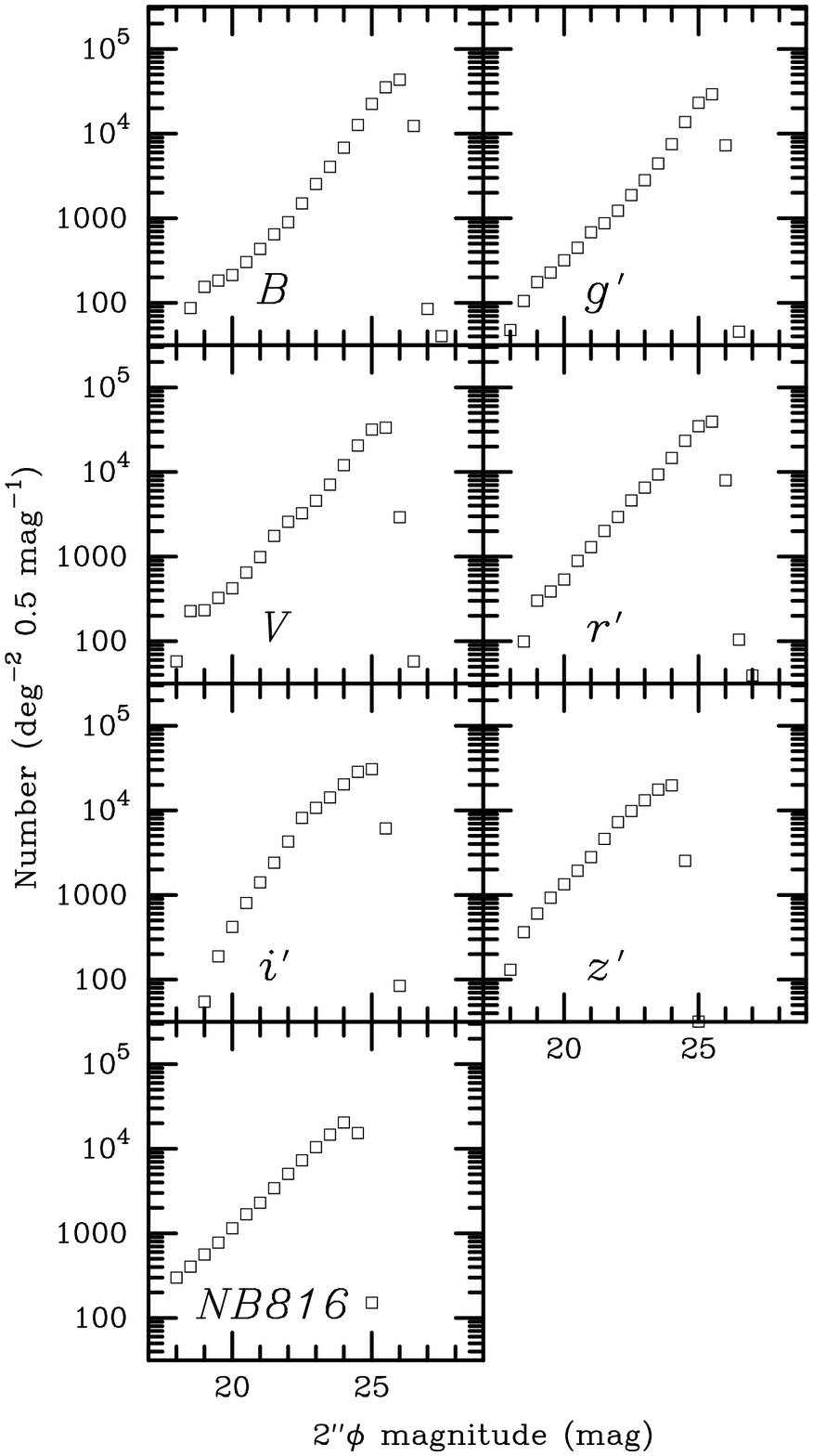}
\caption{Number counts of the COSMOS field as a function of
 $2^{\prime\prime}$ diameter magnitude.
\label{Numc2}}
\end{figure}
\clearpage

\begin{figure}[ht]
\epsscale{0.6}
\plotone{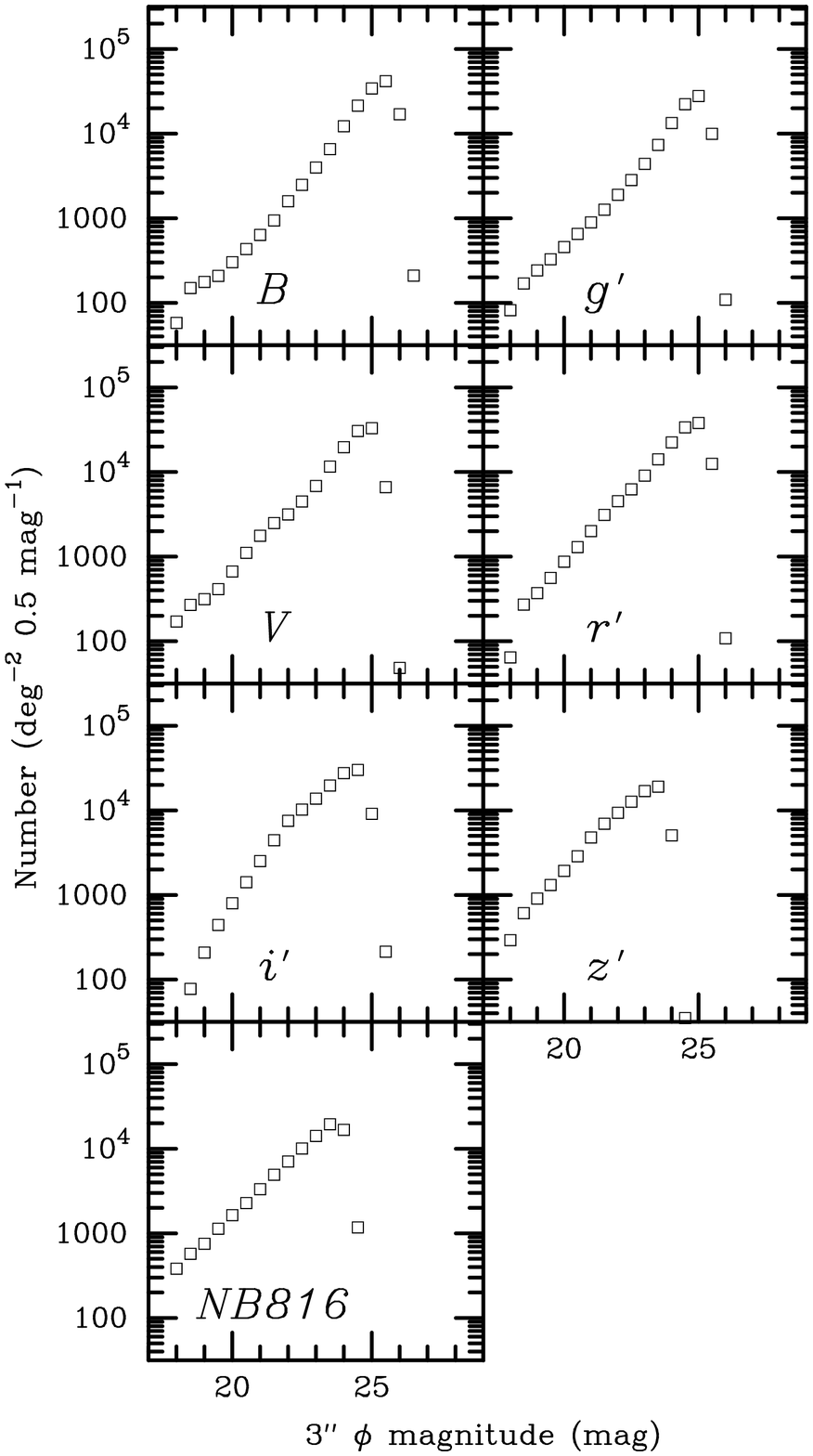}
\caption{Number counts of the COSMOS field as a function of
 $3^{\prime\prime}$ diameter magnitude.
\label{Numc3}}
\end{figure}
\clearpage

\begin{figure}[ht]
\epsscale{1.0}
\plotone{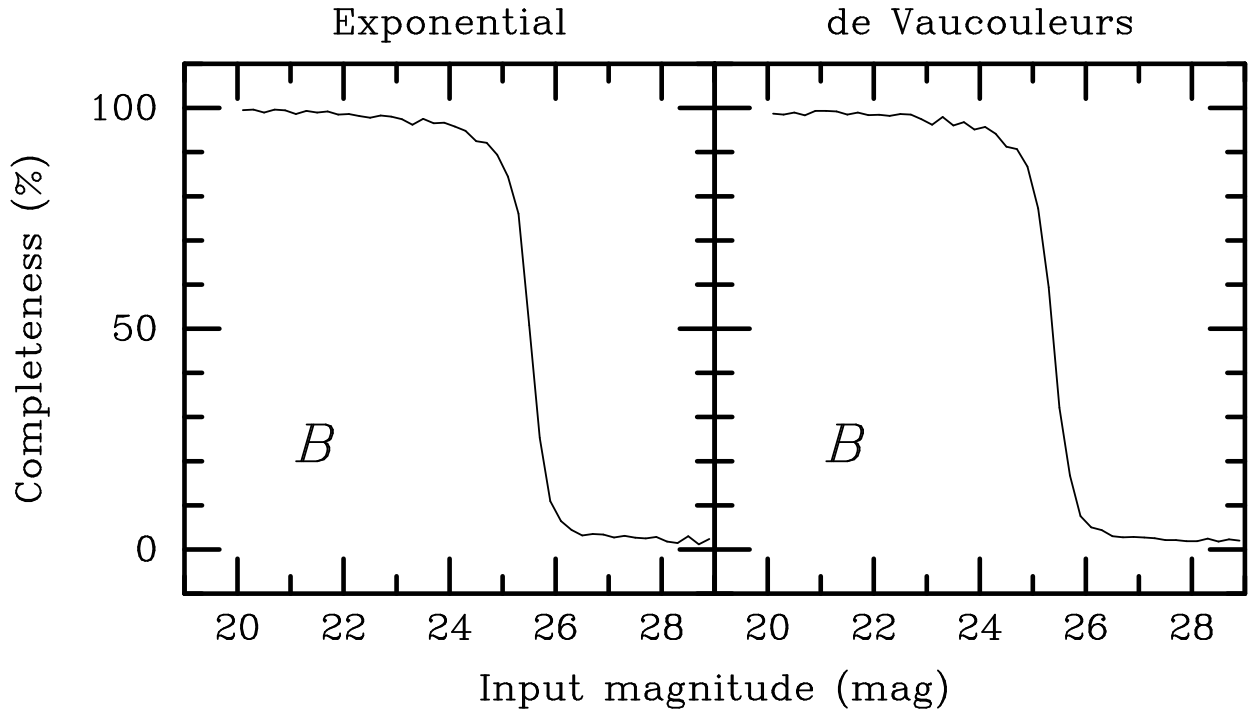}
\caption{Detection completeness estimated by 
simulation in the $B$ band as a function
of input total magnitude of model galaxies.
The left panel shows the case for model galaxies with the light profiles of
the exponential law and the right panel shows the case for those with
de Vaucouleurs' law profile.
\label{Compb}}
\end{figure}
\clearpage

\begin{figure}[ht]
\plotone{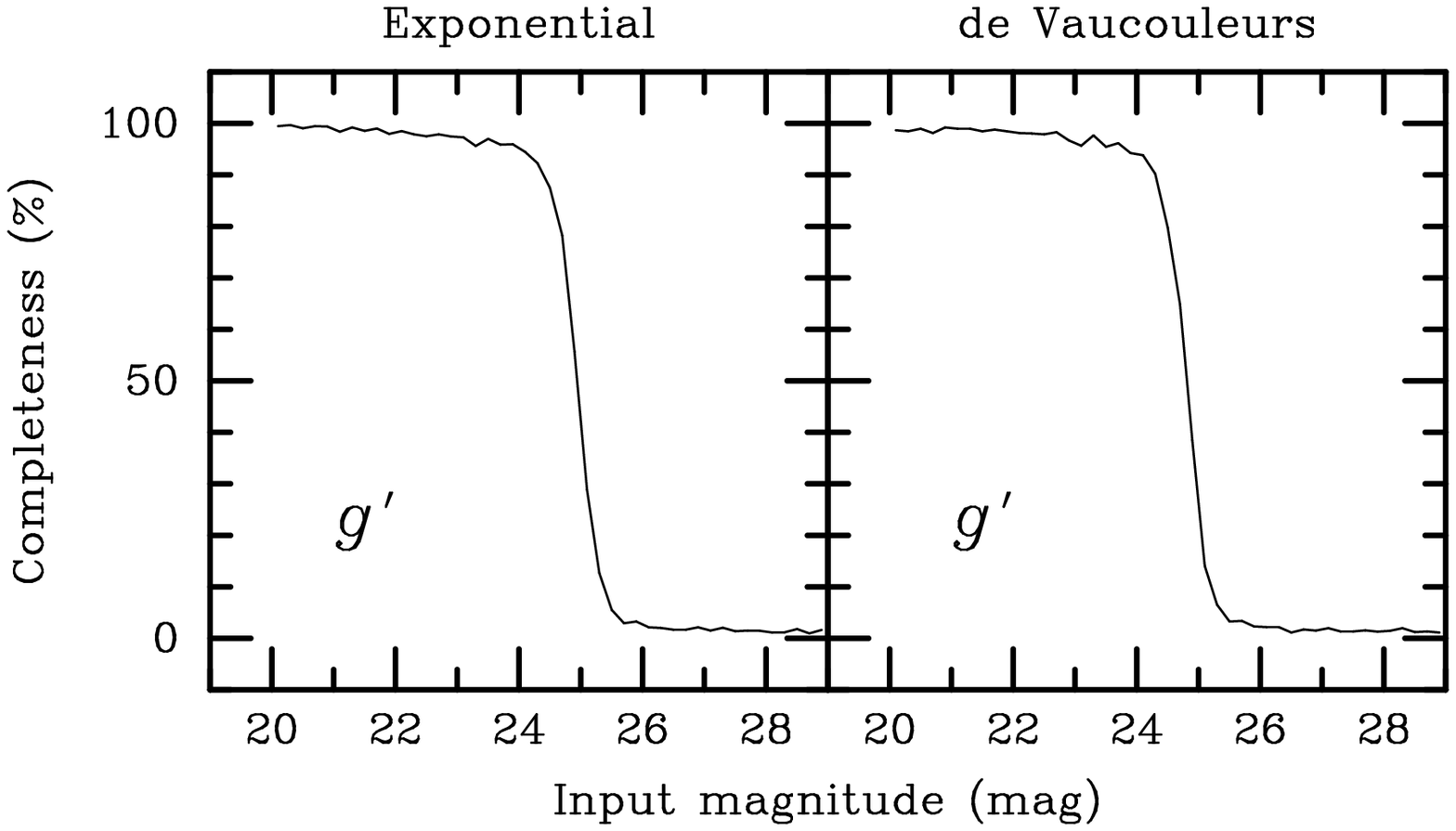}
\caption{The same as Figure \ref{Compb} but for the $g^\prime$ band.
\label{Compgp}}
\end{figure}
\clearpage

\begin{figure}[ht]
\plotone{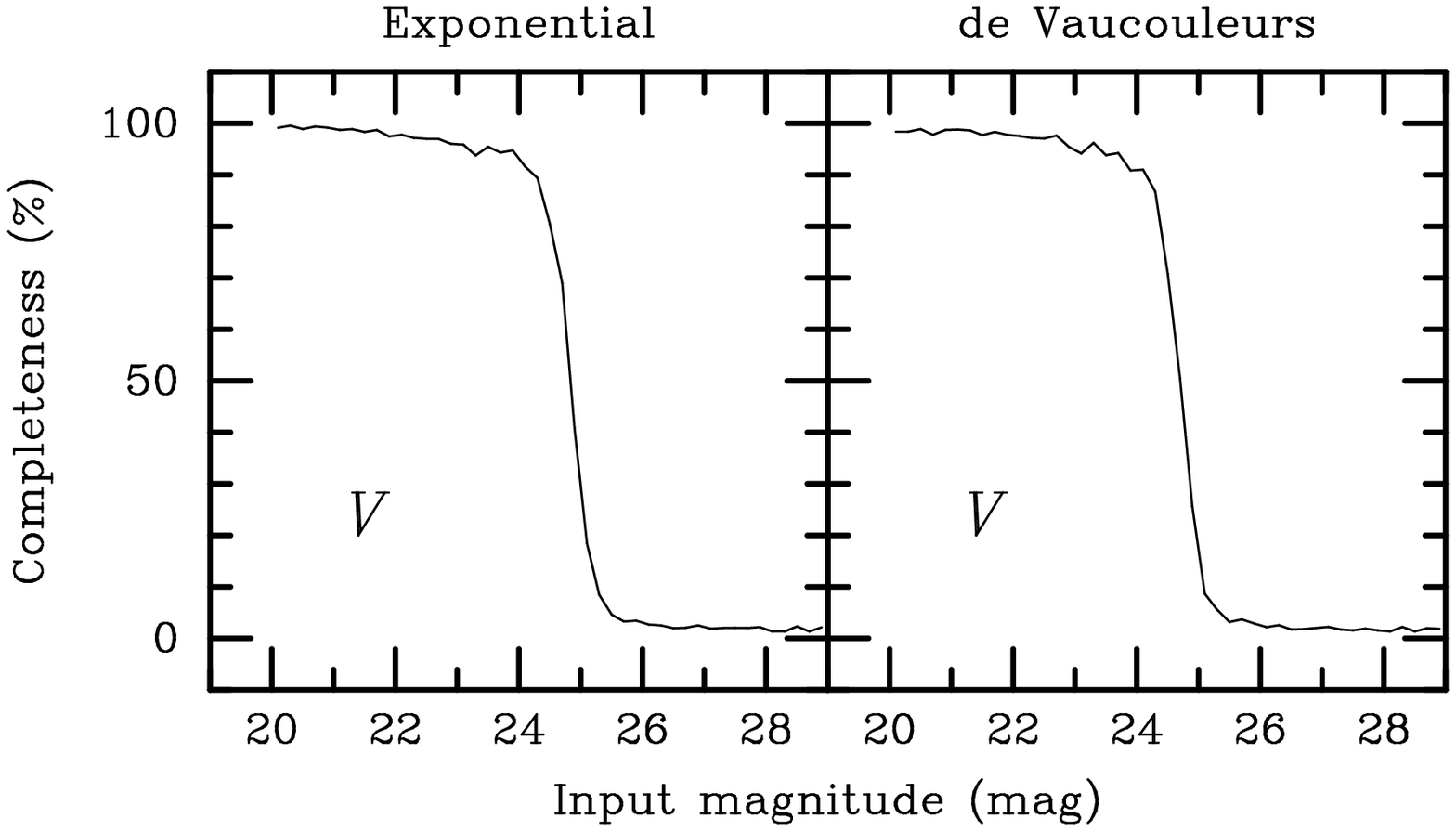}
\caption{The same as Figure \ref{Compb} but for the $V$ band.
\label{Compv}}
\end{figure}
\clearpage

\begin{figure}[ht]
\plotone{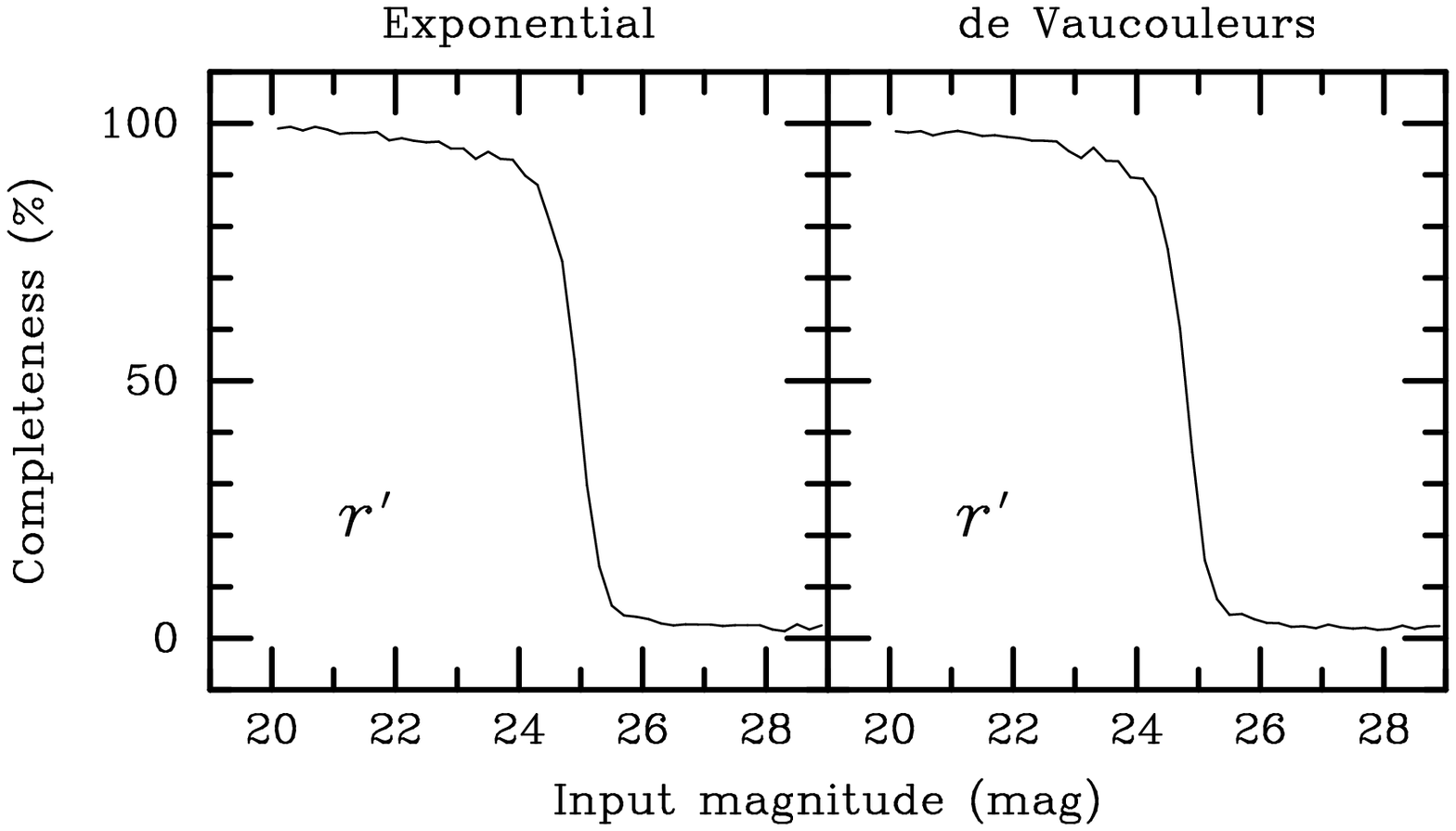}
\caption{The same as Figure \ref{Compb} but for the $r^\prime$ band.
\label{Comprp}}
\end{figure}
\clearpage

\begin{figure}[ht]
\plotone{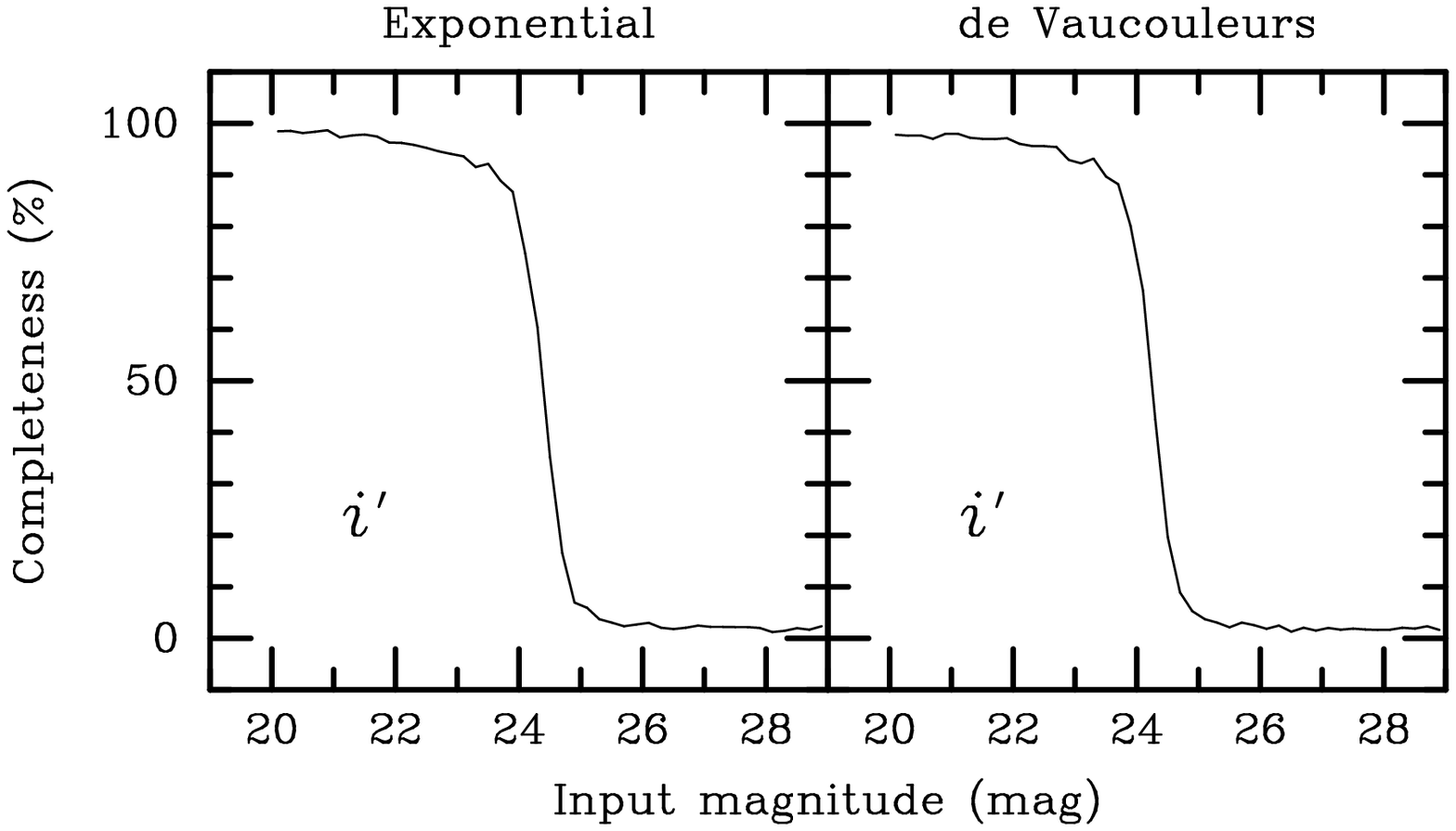}
\caption{The same as Figure \ref{Compb} but for the $i^\prime$ band.
\label{Compip}}
\end{figure}
\clearpage

\begin{figure}[ht]
\plotone{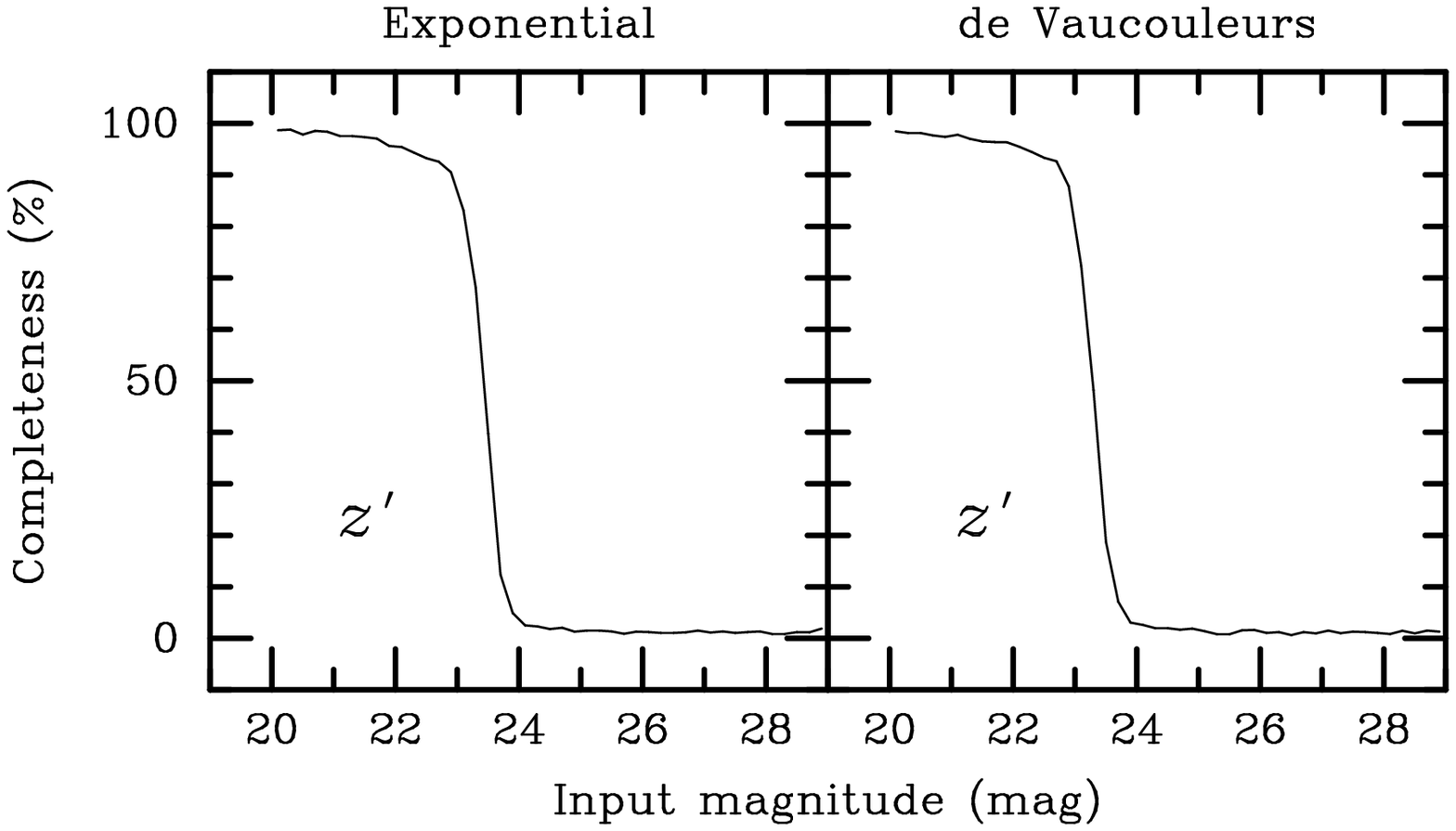}
\caption{The same as Figure \ref{Compb} but for the $z^\prime$ band.
\label{Compzp}}
\end{figure}
\clearpage

\begin{figure}[ht]
\plotone{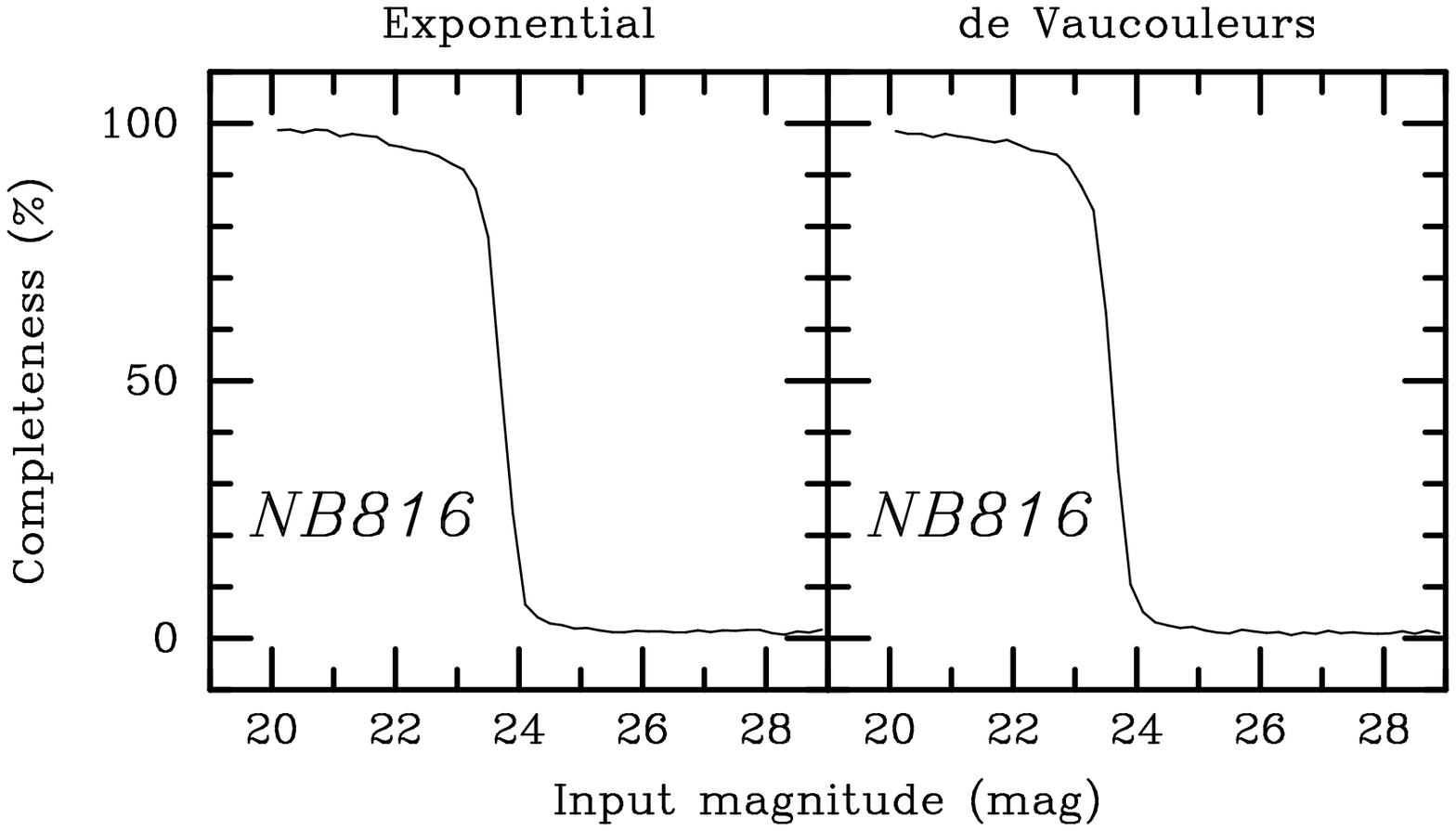}
\caption{The same as Figure \ref{Compb} but for the $NB816$ band.
\label{Compnb}}
\end{figure}
\clearpage

\begin{figure}[ht]
\epsscale{0.6}
\plotone{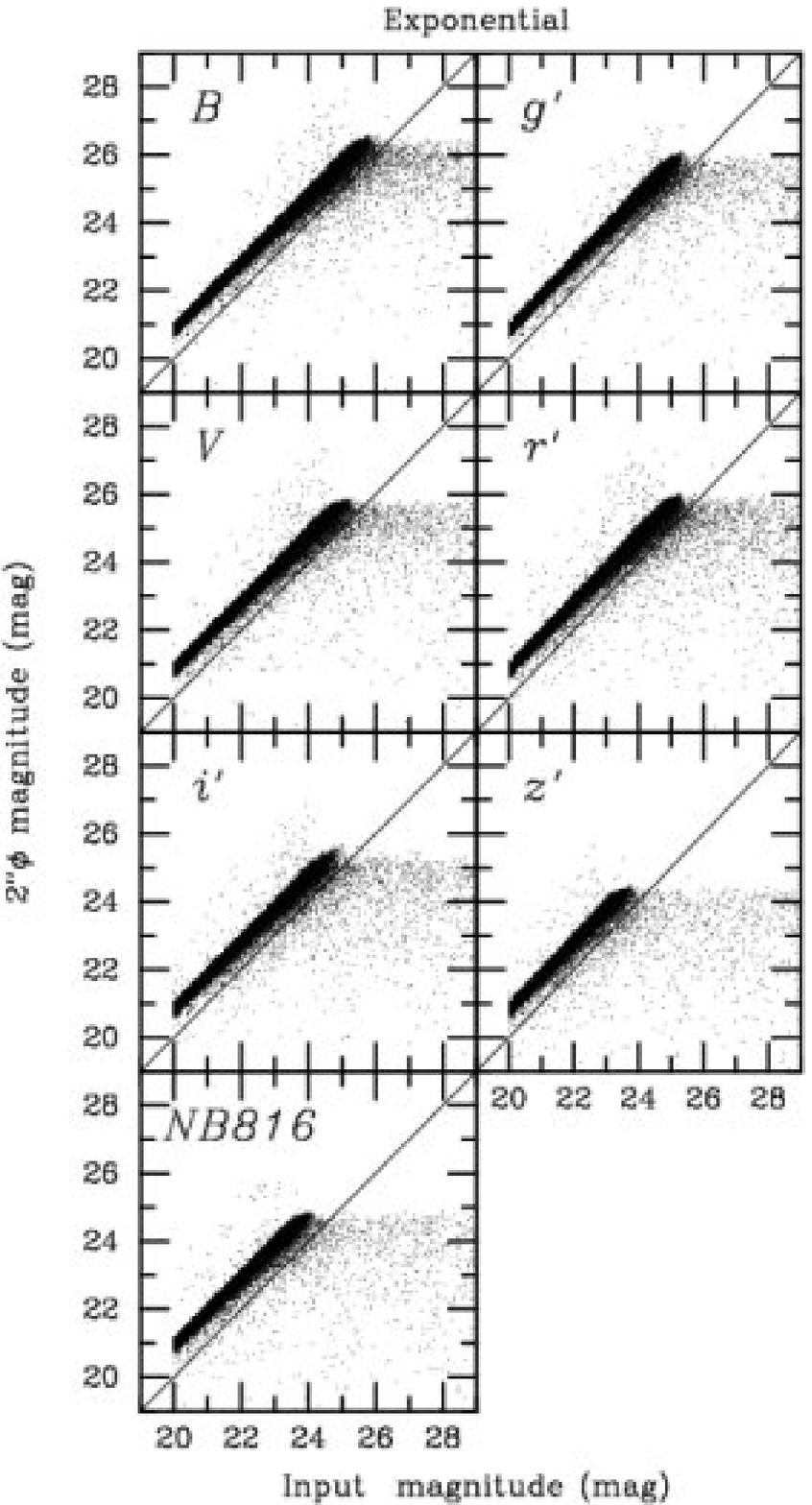}
\caption{The relation between
aperture magnitude of $2^{\prime\prime}$ diameter and input total magnitude
for the detected model galaxies with exponential profiles
used in the completeness analysis.
\label{Maga2exp}}
\end{figure}
\clearpage

\begin{figure}[ht]
\epsscale{0.6}
\plotone{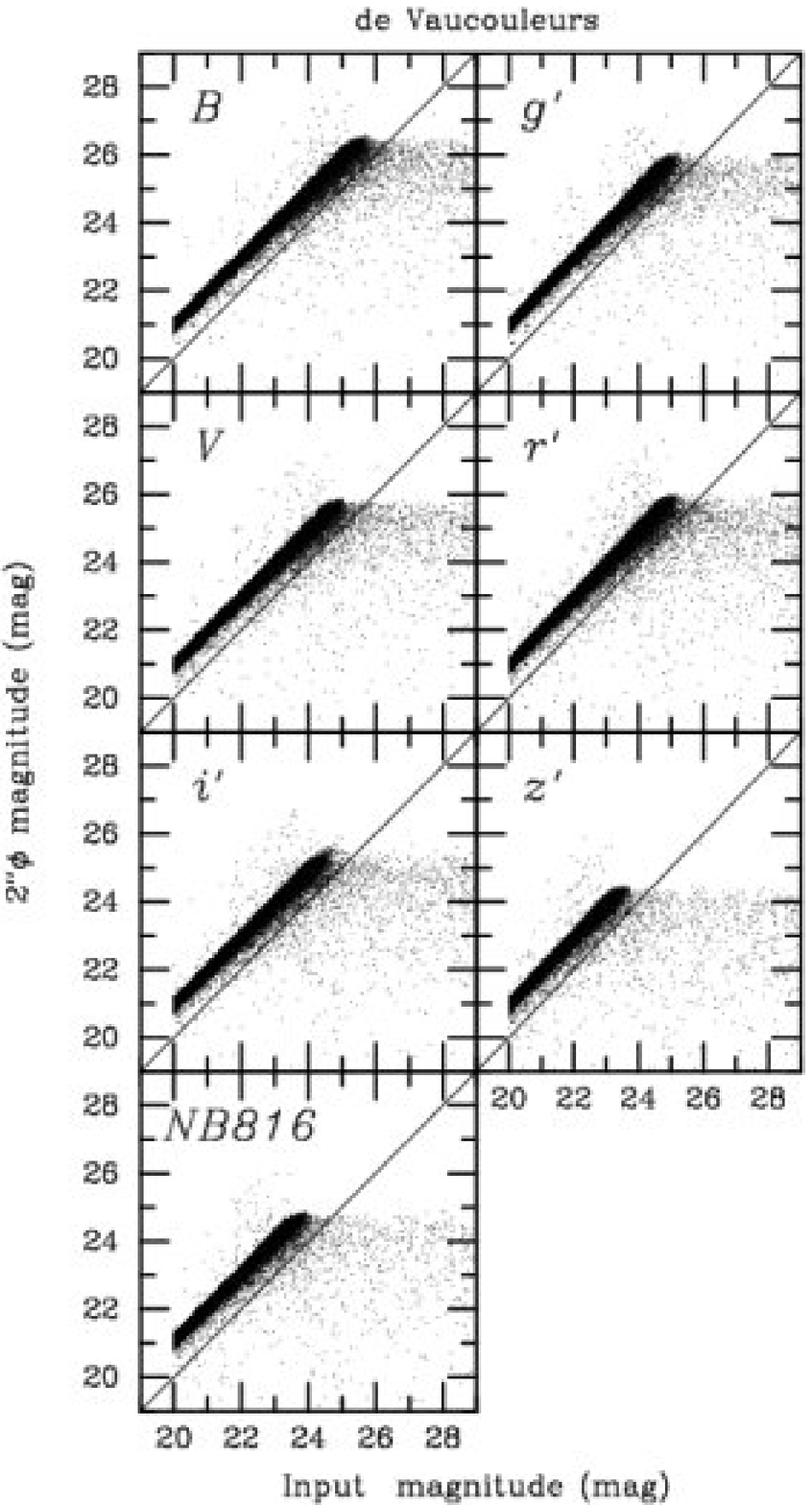}
\caption{The relation between
aperture magnitude of  $2^{\prime\prime}$ diameter and input total magnitude
for the detected model galaxies with de Vaucouleurs' law profiles
used in the completeness analysis.
\label{Maga2devau}}
\end{figure}
\clearpage

\begin{figure}[ht]
\epsscale{0.6}
\plotone{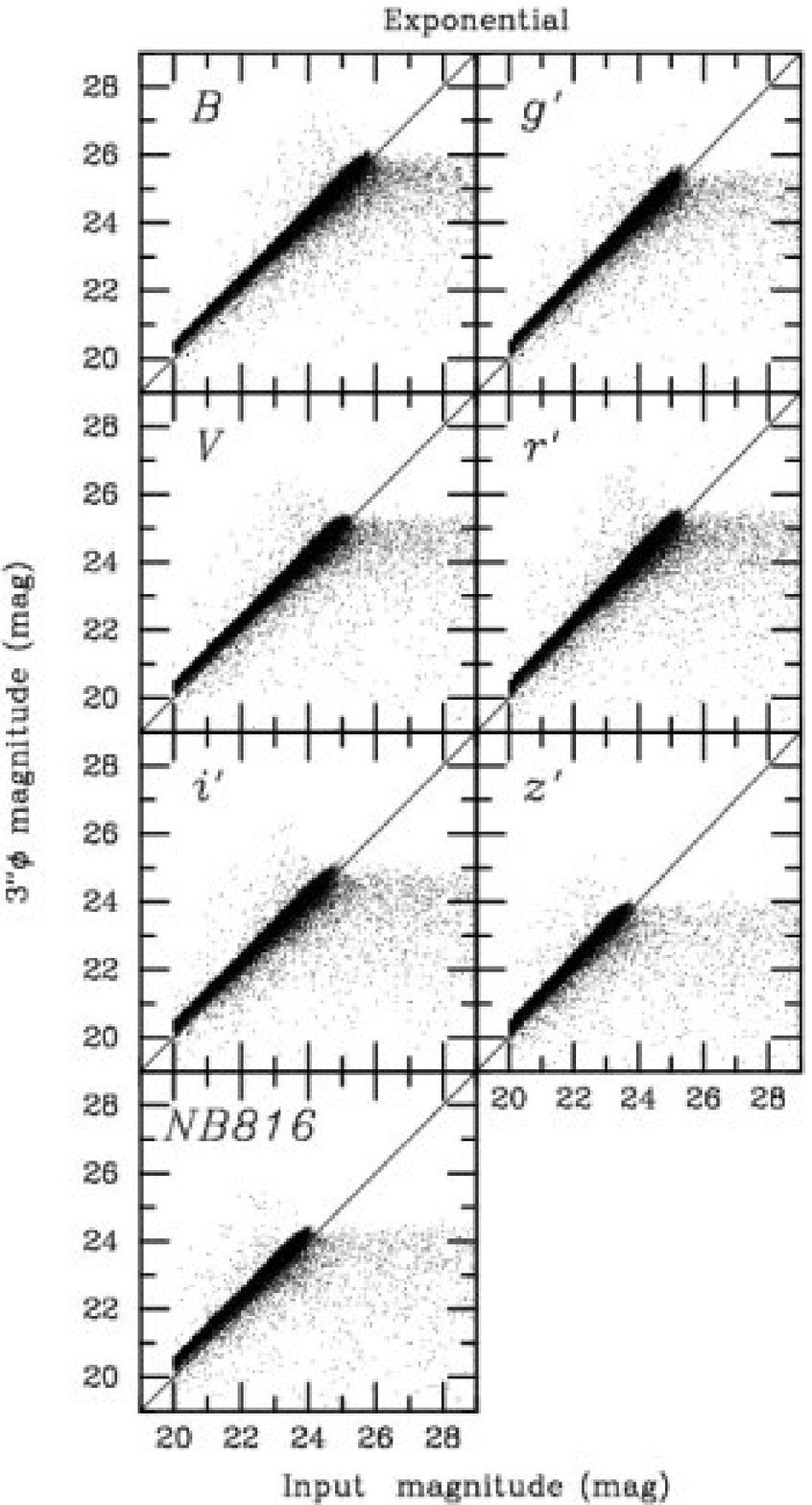}
\caption{The relation between
aperture magnitude of  $3^{\prime\prime}$ diameter and input total magnitude
for the detected model galaxies with exponential profiles
used in the completeness analysis.
\label{Maga3exp}}
\end{figure}
\clearpage

\begin{figure}[ht]
\epsscale{0.6}
\plotone{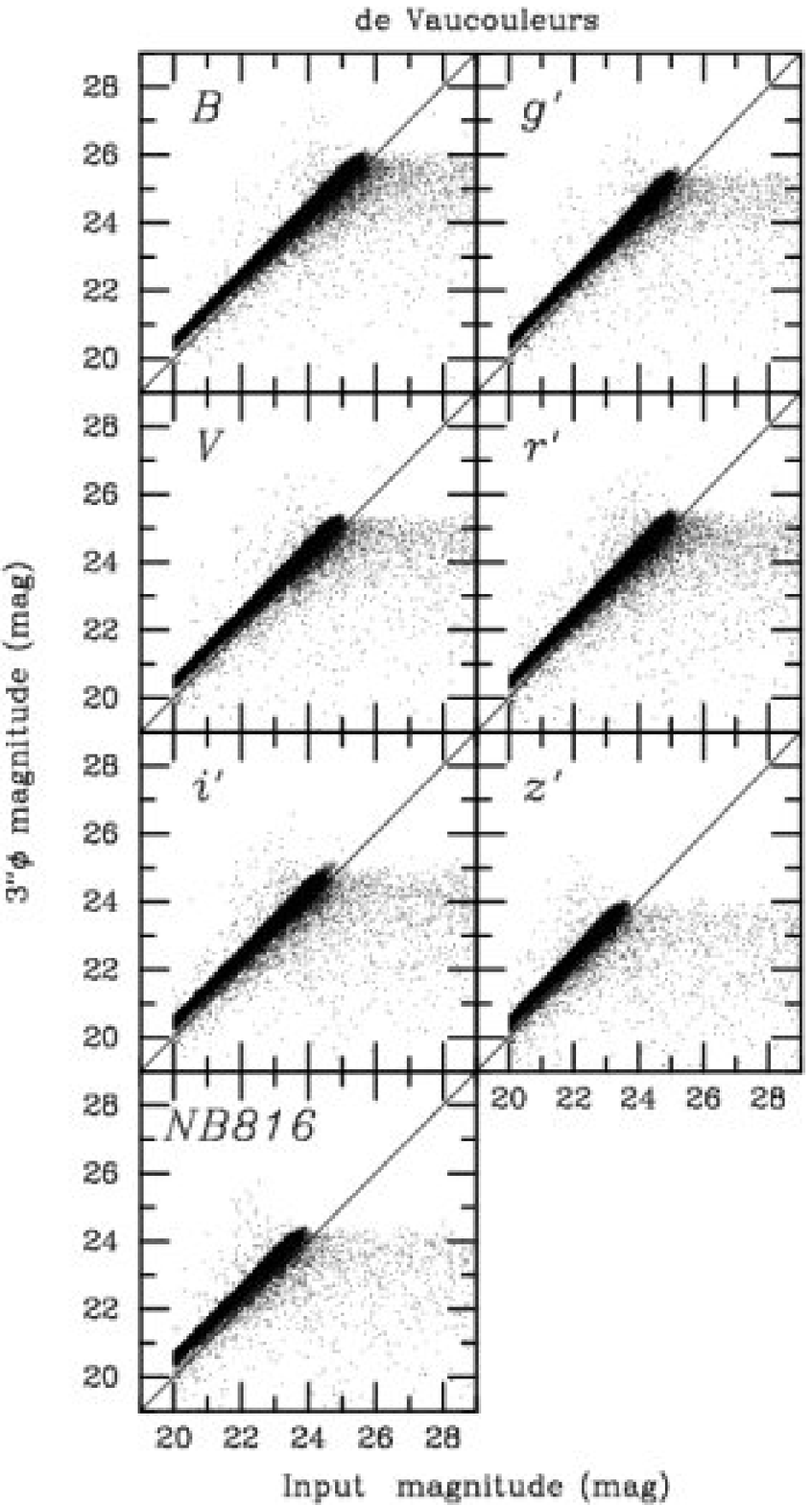}
\caption{The relation between
aperture magnitude of  $3^{\prime\prime}$ diameter and input total magnitude
for the detected model galaxies with de Vaucouleurs' law profiles
used in the completeness analysis.
\label{Maga3devau}}
\end{figure}



\begin{references}
\reference{1}{Aussel, H., et al. 2006, ApJ, submitted}
\reference{1}{Bertin, E., \& Arnouts, S. 1996, A\&AS, 117, 393}
\reference{1}{Capak, P., et al. 2006a, ApJ, submitted}
\reference{1}{Capak, P., et al. 2006b, ApJ, submitted}
\reference{1}{Iye, M., et al. 2004, PASJ, 56, 381}
\reference{1}{Kaifu, N., et al. 2000, PASJ, 52, 1}
\reference{1}{Landolt, A. U. 1992, AJ, 104, 340}
\reference{1}{Miyazaki, S., et al. 2002, PASJ, 54, 833}
\reference{1}{Mobasher, B., et al. 2006, ApJ, submitted}
\reference{1}{Koekemoer, A., et al. 2006, ApJ, submitted}
\reference{1}{Murayama, T., et al. 2006, ApJ, submitted}
\reference{1}{Sasaki, S. S., et al. 2006, ApJ, submitted}
\reference{1}{Sanders, D. B., et al. 2006a, ApJ, submitted}
\reference{1}{Scoville, N. Z., et al. 2006a, ApJ, submitted}
\reference{1}{Scoville, N. Z., et al. 2006b, ApJ, submitted}
\reference{1}{Shioya, Y., et al. 2006, ApJ, submitted}
\reference{1}{Takahashi, M. I., et al. 2006, ApJ, submitted}
\reference{1}{Taniguchi, Y., et al. 2005, JKAS, 38, 187}
\reference{1}{York, D. G., et al. 2000, AJ, 120, 1579}
\end{references}
\end{document}